\documentclass[11pt,a4paper]{article}

\usepackage{epsf,epsfig,psfrag,graphics,color}
\usepackage{amsmath}

\usepackage{geometry}
\usepackage{cite}

\def\CR{\nonumber\\}
\def\calO{\mathcal{O}}
\def\as{\alpha_s}
\def\beq{\begin{eqnarray}}
\def\eeq{\end{eqnarray}}
\def\vareps{\varepsilon}
\def\no{\nonumber}
\def\eps{\epsilon}
\def\om{\omega}
\def\lbar{\bar{\Lambda}}
\def\slash#1{#1 \hskip-0.5em /}

\begin{document}
\thispagestyle{empty}

\begin{flushright}
 TTP08-07\\
 SFB/CPP-08-13\\
 SI-HEP-2007-20 
\end{flushright}

\vspace{\baselineskip}

\begin{center}
\vspace{3\baselineskip}
\textbf{\Large 
Modelling light-cone distribution amplitudes\\[0.5em]
from non-relativistic bound states}\\

\vspace{3\baselineskip}
{\sc G. Bell$^a$ and Th.~Feldmann$^{b,}$\footnote{Also at:
Technische Universit\"at M\"unchen, Physik Department, 85747 Garching,
Germany. }}\\

\vspace{0.7cm}
{\sl ${}^a$ Institut f\"ur Theoretische Teilchenphysik,\\ Universit\"at
  Karlsruhe, D-76128 Karlsruhe, Germany\\}  
\vspace{0.4cm}
{\sl $^b$ Fachbereich Physik, Theoretische Physik I,\\ Universit\"at
  Siegen, Emmy Noether Campus, D-57068 Siegen, Germany} 
\vspace{3\baselineskip}

\vspace*{0.5cm}
\textbf{Abstract}\\
\vspace{1\baselineskip}
\parbox{0.9\textwidth}{
We calculate light-cone distribution amplitudes for non-relativistic
bound states, including radiative corrections from relativistic gluon
exchange to first order in the strong coupling constant. We distinguish
between bound states of quarks with equal (or similar) mass, $m_1\sim
m_2$, and between bound states where the quark masses are hierarchical,
$m_1 \gg m_2$. For both cases we calculate the distribution amplitudes
at the non-relativistic scale and discuss the renormalization-group
evolution for the leading-twist and 2-particle distributions. Our
results apply to hard exclusive reactions with non-relativistic bound
states in the QCD factorization approach like, for instance, $B_c \to
\eta_c \ell \nu$ or $e^{+} e^{-} \to J/\psi \eta_c$. They also serve as
a toy model for light-cone distribution amplitudes of light mesons or
heavy $B$ and $D$ mesons, for which certain model-independent properties
can be derived. In particular, we calculate the anomalous dimension for
the $B$ meson distribution amplitude $\phi_B^-(\om)$ in the Wandzura-Wilczek
approximation and derive the according solution of the evolution
equation at leading logarithmic accuracy.} 
\end{center}

\clearpage

\thispagestyle{empty}
\tableofcontents

\newpage
\setcounter{page}{1}

\section{Introduction}
\label{sec:intro}

Exclusive hadron reactions with large momentum transfer involve strong
interaction dynamics at very different momentum scales. In cases where
the hard-scattering process is dominated by light-like distances, the
long-distance hadronic information is given in terms of so-called
light-cone distribution amplitudes (LCDAs) which are defined from
hadron-to-vacuum matrix elements of non-local operators with quark and
gluon field operators separated along the light-cone
\cite{Braun:1989iv,Ball:1998je} and
\cite{Grozin:1996pq,Beneke:2000wa}. LCDAs appear in the so-called pQCD
approach to hard exclusive reactions
\cite{ERBL,Chernyak:1983ej,Li:1994iu}, in the QCD factorization approach
to heavy-to-light transitions \cite{BBNS}, in soft-collinear effective
theory \cite{SCETa,Beneke:2002ph}, as well as in the light-cone sum rule
approach to exclusive decay amplitudes
\cite{Balitsky:1989ry,Braun:1988qv,Chernyak:1990ag} (for a recent
review, see \cite{Colangelo:2000dp}). 

Representing universal hadronic properties, LCDAs can either be
extracted from experimental data, or they have to be constrained by
non-perturbative methods. The most extensively studied and probably best
understood case is the leading-twist pion LCDA, for which
phenomenological constraints
\cite{Kroll:1996jx,Schmedding:1999ap,Bakulev:2002uc} from the
$\pi-\gamma$ transition form factor \cite{Gronberg:1997fj}, as well as
estimates for the lowest moments from QCD sum rules
\cite{Ball:1998je,Khodjamirian:2004ga,Ball:2006wn} and lattice QCD
\cite{lattice,Gockeler:2005jz} exist. On the other hand, our knowledge
on LCDAs for heavy $B$ mesons
\cite{Grozin:1996pq,Lange:2003ff,Braun:2003wx}, and even more so for
heavy quarkonia \cite{Ma:2006hc,Braguta:2006wr}, had been relatively
poor until recently.  

Although LCDAs, in general, are not calculable in QCD perturbation
theory, their evolution with the factorization scale (which is set by
the momentum transfer of the hard process) can be calculated and is well
understood, both, for light mesons \cite{ERBL} and for heavy mesons
\cite{Lange:2003ff}. The situation becomes somewhat simpler, if the
hadron under consideration can be approximated as a non-relativistic
bound state of two sufficiently heavy quarks. In this case we expect
exclusive matrix elements -- like transition form factors
\cite{Bell:2006tz} and, in particular, the LCDAs --  to be calculable
perturbatively, since the quark masses provide an intrinsic physical
infrared regulator.  

In this article, we are going to calculate the LCDAs for
non-relativistic meson bound states including relativistic QCD
corrections to first order in the strong coupling constant at the
non-relativistic matching scale which is set by the mass of the lighter 
quark in the hadron. We discuss twist-2 and twist-3 LCDAs for 2-particle
Fock states with approximately equal quark masses (for instance an
$\eta_c$ meson), as well as 2-particle and 3-particle LCDAs for heavy
mesons (like the $B_c$), where one of the quark masses is considered to
be much larger than the second one ($m_b \gg m_c$). Our results can also
be viewed as a toy model for possible parameterizations of LCDAs for
relativistic bound states, like the pion, kaon or $B_q$ meson at a low
input scale, which may be evolved to the appropriate higher scales using
the standard renormalization group equations in QCD (or HQET,
respectively).  

Our paper is organized as follows. In the following section we give a
short introduction to the non-relativistic approximation and collect the
definitions and properties of LCDAs for light and heavy mesons. The main
result of our paper, the corrections from relativistic gluon exchange,
are presented in Section~\ref{sec4}. Here we also derive
model-independent results for the $B$ meson LCDAs $\phi_B^+$ and
$\phi_B^-$, namely the cut-off dependence of positive moments and the
anomalous dimension kernels, and investigate the impact of the
3-particle LCDAs to the Wandzura-Wilczek approximation beyond
tree-level. We discuss the effect of QCD evolution above the
non-relativistic matching scale in Section~\ref{sec5}, including a new
result for the $B$ meson LCDA $\phi_B^-$, before we conclude. Some
technical details of the calculation are collected in an appendix. Some 
of our results have already appeared in a proceedings article
\cite{Feldmann:2007id}.

\section{Light-cone distribution amplitudes and the non-relativistic 
  limit}

\subsection{Non-relativistic approximation}

\label{sec:NRapprox}

\begin{figure}[t!]\psfrag{w}{\ }
\centerline{\parbox{0.9\textwidth}{\centerline{\epsfclipon\includegraphics[width=10cm,clip=on]{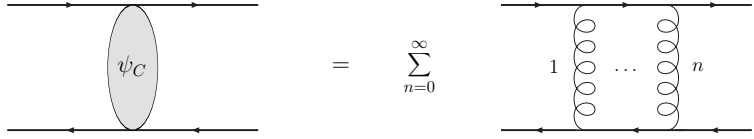}}}}
\vspace{2mm}
\centerline{\parbox{0.9\textwidth}{\caption{\label{fig:CoulombResum}
      {\small Resummation of potential gluons into a non-relativistic
        Coulomb wave-function.}}}} 
\end{figure}

The wave function for a non-relativistic (NR) bound state of a quark and
an antiquark with respective masses $m_1$ and $m_2$ can be obtained from
the resummation of NR (\emph{potential}) gluon exchange as sketched in
Figure~\ref{fig:CoulombResum}. The solution of the corresponding
Schr\"odinger equation with Coulomb potential yields
\begin{align}
\psi_\text{C}(\vec{p}) & {} \propto
\frac{\kappa^{5/2}}{\left(\kappa^2+\left|\vec{p}\,\right|^2\right)^2},
\label{eq:NRwavefunction}
\end{align}
where $\kappa=m_r\alpha_s C_F$ and $m_r=m_1 m_2/(m_1+m_2)$ is the
reduced mass. The normalization of the wave function gives the
(non-relativistic) meson decay constant   
\begin{align}
f_\text{NR} & {} =  \frac{2 \sqrt{N_c}}{\pi} \;
\frac{\kappa^{3/2}}{(m_1+m_2)^{1/2}}. \label{eq:NRdecayconst}
\end{align}
For more details and references to the original literature, see
e.g.~\cite{Brambilla:2004wf} (also \cite{Bell:2006tz}). 

In this approximation, the $\bar{B}_c$ meson is entirely dominated by
the 2-particle Fock state built from a bottom quark with mass $m_1
\equiv M = m_b$ and a charm antiquark with mass $m_2 \equiv m =
m_c$. Consequently to first approximation in the NR expansion, the
$\bar{B}_c$ meson consists of a quark with momentum $M v_\mu$ and an
antiquark with momentum $m v_\mu$, where $v_\mu$ is the four-velocity of
the $\bar{B}_c$ meson ($v^2=1$). The spinor degrees of freedom for the
$\bar{B}_c$ meson are represented by the Dirac projector $\frac12
(1+\slash{v})\gamma_5$. Similarly, a pseudoscalar $\eta_c$ meson is
interpreted as a $c\bar c$ bound state where both constituents have
approximately equal momenta $m v_\mu$. 

The non-relativistic approximation can also serve as a toy model for
bound states of light (relativistic) quarks. We will in the following
refer to ``heavy mesons'' as "$B$" (where we mean the realistic example
of a $B_c$ meson, or the toy model for a $B_q$ meson) and ``light
mesons'' as "$\pi$" (where the realistic example is $\eta_c$, and the
toy-model application would be the pion or also the kaon for $m_1\neq
m_2$).

\subsection{Definition of LCDAs for light pseudoscalar mesons} 

Following~\cite{Braun:1989iv,Ball:1998je} we define the 2-particle LCDAs
of a light pseudoscalar meson via  
\begin{align}
\langle \pi(P) | \bar{q}_1(y) \, [y,x] \, \gamma_\mu \gamma_5 \, q_2(x)
| 0\rangle &{}= -i f_\pi  \int_0^1 \! du \; e^{i (u \, p \cdot y+\bar{u}
  \, p \cdot x)} \; \left[ p_\mu \, \phi_\pi(u) + \frac{m_\pi^2}{2 \,
    p\cdot z}\,z_\mu\, g_\pi(u)\right], 
\CR
\langle \pi(P) | \bar{q}_1(y) \, [y,x] \, i \gamma_5 \, q_2(x) | 0
\rangle &{}= f_\pi \, \mu_\pi  \int_0^1 \! du \; e^{i (u \, p \cdot
  y+\bar{u} \, p \cdot x)} \; \phi_p(u) ,  
\CR
\langle \pi(P) | \bar{q}_1(y) \, [y,x] \, \sigma_{\mu \nu} \gamma_5 \,
q_2(x) |0\rangle &{}= i f_\pi\, \tilde{\mu}_\pi (p_\mu z_\nu - p_\nu
z_\mu) \int_0^1 \! du \; e^{i (u \, p \cdot y+\bar{u} \, p \cdot x)} \;
\frac{\phi_\sigma(u)}{2D-2} 
\label{eq:Pion:def} 
\end{align}
with two light-like vectors $z_\mu=y_\mu-x_\mu$ and
$p_\mu=P_\mu-m_\pi^2/(2 P \, \cdot z)\,z_\mu$, and $u = 1-\bar u$
denoting the light-cone momentum fraction of the quark $q_1$. The gauge
link factor is denoted as 
\begin{align}
[y,x]=\mathcal{P} \exp \left[i g_s \int_0^1 dt \;
(y-x) \cdot A(ty+(1-t)x) \right] \,.
\end{align}
$\phi_\pi(u)$ is the twist-2 LCDA, while $\phi_p(u)$ and
$\phi_\sigma(u)$ are twist-3. For completeness, we have also quoted the
twist-4 LCDA $g_\pi(u)$ which, like the 3-particle LCDAs, will not be
considered further in this work.\footnote{Notice that there are
  additional two-gluon LCDA for flavour singlet mesons which we will not
  consider here, because in the non-relativistic limit glueballs
  decouple from the $q\bar q$ states and the 2-gluon LCDA is only
  generated by higher-order relativistic corrections. For the definition
  of the 3-particle LCDAs, see~\cite{Braun:1989iv,Ball:1998je}.} All
LCDAs are normalized to 1, such that the prefactors in
(\ref{eq:Pion:def}) are defined in the local limit $x \to y$. In the
definition of $\phi_\sigma(u)$, we have included a factor $3/(D-1)$,
such that the relation between $\mu_\pi$ and $\tilde \mu_\pi$ from the
equations of motion (see below) is maintained in $D\neq 4$ dimensions.

\subsubsection{Equations of motion}

The equations of motion (eom) provide relations between the matrix
elements defined in (\ref{eq:Pion:def}). Following~\cite{Ball:1998je} we
obtain  
\begin{align}
\frac{m_\pi^2}{2} \Big[ \phi_\pi(u) + g_\pi(u) \Big] 
&{}= (m_1+m_2) \, \mu_\pi \, \phi_p(u) + \ldots, 
\no \\  
\mu_\pi \, \phi_p(u) + \frac{\tilde{\mu}_\pi}{D-1} 
\left[ (2-D) \,\phi_\sigma(u) + \frac{2u-1}{2} \,\phi_\sigma'(u) \right]
 &{}= (m_1 + m_2) \, \phi_\pi(u) + \ldots, 
\no \\ 
(2u-1) \,\mu_\pi \, \phi_p(u) + \frac{\tilde{\mu}_\pi}{2D-2}
\,\phi_\sigma'(u) &{}=  (m_1-m_2) \,\phi_\pi(u) +\ldots,
\label{eq:eom:non-local}
\end{align}
where the ellipsis denote contributions from 3-particle LCDAs which we
do not specify here. In the local limit the contributions from the
3-particle LCDAs drop out and integration of (\ref{eq:eom:non-local})
yields 
\begin{align} 
\mu_\pi &{}= \frac{m_\pi^2}{m_1 + m_2}, 
\qquad
\tilde{\mu}_\pi = \mu_\pi - (m_1 + m_2)
\label{eq:eom:local1}
\end{align}
and
\begin{align}
\int_0^1 du \, u \; \phi_p(u)&{}= \frac12 + \frac{m_1-m_2}{2\mu_\pi}\,.
\label{eq:eom:local2}
\end{align}
Notice that the relations
(\ref{eq:eom:non-local},\ref{eq:eom:local1},\ref{eq:eom:local2}) hold
for the \emph{bare} (unrenormalized) parameters and distribution
amplitudes.

\subsubsection{Tree-level result}

At tree level, and in leading order of the expansion in the relative
velocities, the quark and the antiquark in the NR wave function simply
share the momentum of the meson according to their masses, $p_i^\mu
\simeq m_i/(m_1+m_2) P^\mu$. For "light" mesons this implies\footnote{
  This behaviour can also be obtained from the ``dense medium limit'' in
  the instanton model \cite{Anikin:1999cx}.} 
\begin{align}
  \phi_\pi(u) &{} \simeq \phi_p(u)  \simeq g_\pi(u) \simeq \delta(u-
  u_0) \,, 
\label{delta}
\end{align}
with $u_0=m_1/(m_1+m_2)$ and $\bar u_0 = m_2/(m_1+m_2)$. Consequently,
all positive and negative moments of the distribution amplitudes are
simply given in terms of the corresponding power of $u_0$. In
particular, the Gegenbauer coefficients are given by 
\begin{align}
  a_n & {} = \frac{2 (2n+3)}{3(2+n)(1+n)} \int_0^1 du \,
  \phi_\pi(u) \, C_n^{(3/2)}(2u-1) 
  \quad\to\quad
  \frac{2 (2n+3)}{3(2+n)(1+n)} \, C_n^{(3/2)}(2u_0-1) \,.
\label{proj}
\end{align}
Notice that $\tilde \mu_\pi \simeq 0$ at tree-level and the
corresponding LCDA $\phi_\sigma(u)$ can only be determined by
considering the corresponding one-loop expressions (see below). The
tree-level solutions (\ref{delta}) fulfill the eom-constraints from
(\ref{eq:eom:non-local}).

\subsection{Definition of LCDAs for heavy pseudoscalar mesons} 

We define the 2-particle LCDAs of a heavy pseudoscalar $B$ meson
following \cite{Grozin:1996pq,Beneke:2000wa},
\begin{align}
\langle 0 | (\bar{q})_\beta(z) \, [z,0] \, (h_v)_\alpha(0) | B( M v) \rangle
 & {} = - \frac{i \hat{f}_B(\mu) M}{4} \left[ \frac{1
+\slash{v}}{2} \left\{ 2 \tilde{\phi}_B^+(t) +
\frac{\tilde{\phi}_B^-(t)-\tilde{\phi}_B^+(t)}{t} \slash{z}
\right\} \gamma_5 \right]_{\alpha \beta} \,,
\label{eq:NRdefLCDABc}
\end{align}
where $v^\mu$ is the heavy meson's velocity, $t\equiv v \cdot z$ and
$z^2=0$. Here $\hat{f}_B$ is the (renormalization-scale dependent) decay
constant in HQET. The Fourier-transformed expressions, which usually
appear in factorization formulas, are given through 
\begin{align}
\tilde{\phi}_B^\pm(t) & {}= \int_0^\infty \! d\omega \; e^{-i\omega t}
\phi_B^\pm(\omega)\,,
\end{align}
where $\omega$ denotes the light-cone energy of the light quark in the
$B$ meson rest frame.

\subsubsection{Equations of motion}

The equations of motion again provide relations between different
LCDAs. Including the effect of the 3-particle LCDAs $\Psi_A, \Psi_V,
X_A, Y_A$ as defined in~\cite{Kawamura:2001jm} (see also
\cite{Grozin:2005iz,Huang:2005kk}), we derive
\begin{align}
& \om\, \phi_B^-(\om)- m\, \phi_B^+(\om) + \frac{D-2}{2} 
\int_0^\om d\eta \; \left[\phi_B^+(\eta)-\phi_B^-(\eta)\right] \CR
& \qquad  =\;
(D-2) \int_0^\om d\eta  \; \int_{\om-\eta}^\infty \frac{d\xi}{\xi} \;
\frac{\partial}{\partial \xi} \; 
\left[ \Psi_A(\eta,\xi)-\Psi_V(\eta,\xi)\right] \,.
\label{eom1}
 \end{align}
The relation (\ref{eom1}) is trivially fulfilled at tree-level and we
will show below that it also holds after including the $\alpha_s$
corrections to the NR limit. In \cite{Kawamura:2001jm}, Kawamura et
al.~discuss a second relation which in the massive case reads  
\begin{align}
 & (\om+m) \, \phi_B^-(\om) + (\om - 2 \lbar -  m) \, \phi_B^+(\om) \CR
 & \qquad \stackrel{?}{=} \;
 -2 \frac{d}{d\om} \int_0^\om d\eta \; \int_{\om-\eta}^\infty 
 \frac{d\xi}{\xi} \; \left[ \Psi_A(\eta,\xi)+X_A(\eta,\xi)\right]
 -2(D-2) \int_0^\om d\eta \; \int_{\om-\eta}^\infty 
 \frac{d\xi}{\xi} \; \frac{\partial \Psi_V(\eta,\xi)}{\partial \xi} \,,
\label{eom2}
\end{align}
with $\lbar=M_B-m_b$. We will show below that the equation (\ref{eom2})
\emph{does not} hold beyond tree level, since the integral on the
right-hand side involving our result for the 3-particle LCDA $X_A$ does
not converge. This confirms the criticism raised in
\cite{Braun:2003wx,Braun:talk} that (\ref{eom2}) is not consistent,
since the renormalization prescription of light-cone operators in HQET
and the expansion into local operators do not commute. Notice that in
contrast to (\ref{eom1}), the derivation of (\ref{eom2}) involves
derivatives with respect to $z^2\neq0$.   

If one neglects the 3-particle distribution amplitudes in
(\ref{eom1}), one arrives at the so-called Wandzura-Wilczek relation
which has first been discussed for a massless light quark in
\cite{Beneke:2000wa}. The generalization to the massive case reads    
\begin{align}
\label{ww0}
\int_0^{\om} d\eta \,\left[\phi_B^-(\eta)-
\phi_B^+(\eta)\right] & {} \simeq \frac{2}{D-2} \left[ \omega
\,\phi_B^-(\omega) - m \, \phi_B^+(\omega) \right] \,,
\end{align}
which again holds for the bare parameters and LCDAs in $D\neq 4$ dimensions.

\subsubsection{Tree-level result}

By the same arguments as for light mesons, at tree-level the quark and
the antiquark in a heavy meson just share the total momentum according
to their masses, such that $\omega = m$. In the NR limit, the 2-particle
LCDAs of a ''heavy'' meson are thus given by 
\begin{align}
  \phi_B^+(\omega) & {} \simeq \phi_B^-(\omega) \simeq \delta(\omega -
  m) \,. 
\label{heavy-tree}
\end{align}
Moreover, at tree level, the moments of the heavy meson LCDAs can be
related to matrix elements of local operators in HQET
\cite{Grozin:1996pq}. The zeroth moment  
$
  \langle 0 | \bar q \, \gamma_5 \, h_v |B \rangle = - i \hat f_B M
$
determines the tree-level normalization of the distribution amplitudes
$\tilde \phi_B^\pm(t=0)\simeq 1$. 
For the first moment, one has the general decomposition 
\begin{align}
 \langle 0| (\bar q)_\beta \, i \overset{\leftarrow\;}{D^\mu} \,
 (h_v)_\alpha|B(v)\rangle & {}  \simeq  - \frac{i M \hat f_B}{4}
 \,\left[  \left( a v^\mu + b \gamma^\mu \right)  (1 +\slash{v})
   \,\gamma_5 \right]_{\alpha \beta} \,. 
\end{align}
Multiplying by $(\gamma_5 \gamma_\mu)_{\beta\alpha}$ and taking into
account the finite light quark mass in the NR set-up, the equation of
motion for the light quark implies $a + 4b = m$. The equation of motion
for the heavy quark is obtained by multiplying with $v^\mu$, from which
one obtains $a + b = \bar \Lambda$, independent of the light-quark
mass. This implies    
$$
a=\frac{4 \bar \Lambda - m}{3}\,, \qquad
b=- \frac{\bar \Lambda - m}{3}\,.
$$
From this we can read off the first moments at tree-level
\begin{align}
\langle \omega \rangle_+ & {} \simeq \frac{i}{\hat f_B M} \, \langle 0|
\bar q \, \gamma_5 \, \slash n_- \, (i n_- \overset{\leftarrow}{D}) \,
h_v|B\rangle = a = \frac{4 \bar \Lambda - m}{3} \,,
\\[0.2em]
\langle \omega \rangle_- & {} \simeq \frac{i}{\hat f_B M} \,  \langle 0|
\bar q \, \gamma_5 \, \slash n_+ \, (i n_- \overset{\leftarrow}{D}) \,
h_v|B\rangle = 2b+a = \frac{2 \bar\Lambda+m}{3}\,,
\end{align}
where we introduced the light-like vectors $n_-^\mu = z^\mu/t$ and
$n_+^\mu =2 v^\mu - n_-^\mu$. In the non-relativistic limit $\bar
\Lambda = m$, and we obtain 
$$
  \langle \omega \rangle_\pm \simeq m  \,.
$$
Notice that the light-quark mass drops out in the sum
$$
   \langle \omega \rangle_+ + \langle \omega \rangle_- = 2\bar \Lambda \,.
$$
We stress that the relation between moments of $\phi_B^\pm(\omega)$ and
local matrix elements in HQET does not hold beyond the tree-level
approximation \cite{Lange:2003ff,Braun:2003wx,Lee:2005gz}.

\section{Relativistic corrections at one-loop} 

\label{sec4}

\begin{figure}[t!pbt]
\linespread{1.1}
\begin{center}
\centerline{\includegraphics[width=13cm]{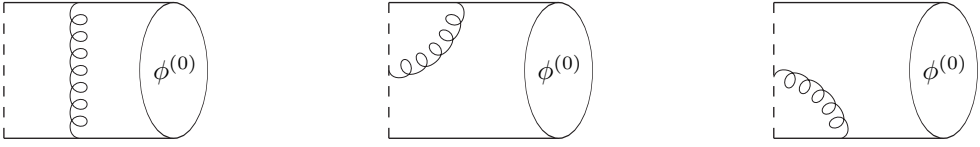}}
\vspace{2mm}
\centerline{\parbox{0.9\textwidth}{ \caption{\label{fig:relcorr} {\small
        Relativistic corrections to the light-cone distribution
        amplitudes. The dashed line indicates the Wilson line in the
        definition of the LCDAs.}}}}  
\end{center}
\end{figure}

The NR bound states are described by parton configurations with fixed
momenta. Relativistic gluon exchange as in Figure~\ref{fig:relcorr} leads
to modifications: First, there is a correction from matching QCD (or, in
the case of heavy mesons, the corresponding low-energy effective theory
HQET) on the NR theory. Secondly, there is the usual evolution under the
change of the renormalization scale \cite{ERBL,Lange:2003ff}. In
particular, the support region for the parton momenta is extended to
$0\leq u\leq1$ for light mesons and $0 \leq \omega < \infty$ for heavy
mesons. In this section we collect the results for LCDAs for ``light'' and
"heavy" mesons including the first-order matching corrections from
relativistic gluon exchange   
\begin{align}
 \phi_M &{}  = \phi_M^{(0)} + \frac{\alpha_s C_F}{4\pi} \, \phi_M^{(1)} + 
{\cal O}(\alpha_s^2) \,.
\end{align}

\subsection{Light mesons}

\subsubsection{Local matrix elements}

We first consider the leading-order relativistic corrections to the
local matrix elements which are given by the vertex-correction and the
wave-function renormalization of the quark fields. We find
\begin{align}
f_\pi &{}= f_\pi^\text{NR}
\left[ 1 + \frac{\as C_F}{4\pi} \left( - 6 + 3 \ \frac{m_1-m_2}{m_1+m_2}
    \, \ln \frac{m_1}{m_2} \right) + \calO(\as^2) \right]
\label{eq:fpires}
\end{align}
and
\begin{align}
\mu_\pi = \frac{m_\pi^2}{
 Z_{m_1}^\text{os} \, m_1^\text{os} +Z_{m_2}^\text{os} \,m_2^\text{os}}
&{}= m_\pi 
 \left[ 1+\frac{\as C_F}{4\pi} \left(
    \frac{3}{\vareps} + 3 \, \ln \frac{\mu^2}{m_1m_2} 
- 3 \ \frac{m_1-m_2}{m_1+m_2} \, \ln \frac{m_1}{m_2} + 4 \right)
+\calO(\as^2) \right],
\no \\
\tilde{\mu}_\pi = \mu_\pi - \frac{m_\pi^2}{\mu_\pi}
 &{}= m_\pi \left[ \frac{\as C_F}{4\pi} \left(
    \frac{6}{\vareps} + 6 \ln \frac{\mu^2}{m_1m_2} 
- 6 \ \frac{m_1-m_2}{m_1+m_2} \, \ln \frac{m_1}{m_2} + 8 \right)
  +\calO(\as^2)\right],
\end{align}
where $m_\pi \simeq m_1^\text{os}+m_2^\text{os}$ in the on-shell
scheme. Our result for the decay constant is in agreement
with~\cite{Braaten:1995ej} and the results for $\mu_\pi$ and
$\tilde{\mu}_\pi$ are consistent with the eom-constraints in
(\ref{eq:eom:local1}).

\subsubsection{The twist-2 LCDA $\phi_\pi(u)$}

Let us start with the case of equal quark masses, e.g.\ in case of a
non-relativistic $\eta_c$ bound state, which may also serve as a
toy-model for the pion LCDA.\footnote{We should keep in mind, however,
  that typically non-perturbative analyses from lattice QCD and sum
  rules find pion distribution amplitudes that are \emph{broader} than
  the asymptotic one, while the non-relativistic model assumes LCDAs
  which are \emph{narrower}. Therefore the application of the toy model
  to the very pion case should not be taken too seriously.}  
 
The first-order relativistic corrections arise from the collinear gluon
exchange diagrams in Figure~\ref{fig:relcorr}, where we also have to
take into account the wave-function renormalization of the external
quark lines (see Appendix~\ref{app:phipi} for details). The local limit
of the light-cone matrix element (\ref{eq:Pion:def}) determines the
relativistic corrections to the NR decay constant (\ref{eq:fpires}) (in
this case, the diagrams with the gluon attached to the Wilson-line do
not contribute). The remaining contributions to the NLO correction for
the leading-twist LCDA contain an UV-divergent piece,   
\begin{align}
\phi_\pi^{(1)}(u) \big|_{\rm div.} &{}= \frac{2}{\vareps} \, \int_0^1 dv
\, V(u,v) \, \phi^{(0)}(v)\,, 
\end{align}
which involves the well-known Brodsky-Lepage evolution kernel
\cite{ERBL}, 
\begin{align}
V(u,v) &{}= \left[ \left(1 + \frac{1}{v-u} \right) \, \frac{u}{v}\
\theta(v-u) + \left(1 + \frac{1}{\bar v-\bar u} \right) \, \frac{\bar
u}{\bar v} \ \theta(u-v)
 \right]_+ \,.
\label{ERBL}
\end{align}
The finite terms after $\overline{\rm MS}$\/-subtraction read 
\begin{align}
\phi_\pi^{(1)}(u;\mu) &{}= 4  \left\{
\left( \ln \frac{\mu^2}{m_\pi^2 \, (1/2-u)^2} -1 \right)\!
 \left[ \left( 1+\frac{1}{1/2-u} \right) u\;\theta(1/2-u) + 
 (u \leftrightarrow \bar u) \right] \right\}_{+} \no \\
& \quad 
+4  \left\{ \frac{u (1-u)}{(1/2-u)^2} 
\right\}_{++}
 \label{eq:NRLCDABNLO} \,.
\end{align}
Here the plus-distributions are defined as 
\begin{align} \label{eq:plusdef}
\int_0^1 \! du \; \big\{\ldots\big\}_{+} \;f(u) & {} \equiv \int_0^1 \!
du \; \big\{\ldots\big\} \; \bigg( f(u) - f(1/2) \bigg) \,,
\\[0.2em]
\int_0^1 \! du \; \big\{\ldots\big\}_{++} \;f(u) &{} \equiv \int_0^1 \!
du \; \big\{\ldots\big\} \; \bigg( f(u) - f(1/2) -f'(1/2) \, (u-1/2)\bigg)\,.
\label{eq:plusplusdef}
\end{align}
From this it follows that
$$
  \int_0^1 du \, \phi^{(1)}_\pi(u;\mu) 
=  \int_0^1 du \, u\, \phi^{(1)}_\pi(u;\mu) = 0 \,, 
$$
such that  the general normalization conditions $\int_0^1 du \,
\phi_\pi(u) = 1$ and $\int_0^1 du \, u\, \phi_\pi(u) = 1/2$ are not
changed. Furthermore, our result for the distribution amplitude obeys
the evolution equation
\begin{align}
\frac{d}{d \ln \mu} \; \phi_\pi(u;\mu) &{}=  \frac{\alpha_s C_F}{\pi}
\int_0^1 dv \; V(u,v) \; \phi_\pi(v;\mu) + {\cal O}(\alpha_s^2) \,.
\label{EE:phipi}
\end{align}
An independent calculation of the leading-twist LCDAs for the $\eta_c$
and $J/\psi$ meson has been presented in~\cite{Ma:2006hc}. Our result is
not in complete agreement with these findings. In particular, we find
that the LCDA quoted in~\cite{Ma:2006hc} is not normalized to unity as
it should be.  

On the other hand, at the non-relativistic scale $\mu \simeq m$, the
distribution amplitude shows a singular behaviour at $u\simeq u_0 =
1/2$. As a consequence, the convergence of the Gegenbauer expansion is
not very good at the non-relativistic scale, with the Gegenbauer
coefficients $a_n$ in (\ref{proj}) only falling off as $1/\sqrt{n}$ (and
alternating signs). A better characterization of the LCDA at $\mu \simeq
m$ is given in terms of the moments 
\begin{align}
 \langle \xi^n \rangle_\pi (\mu)\equiv &
\int_0^1 du \, (2u-1)^n \, \phi_\pi(u;\mu) \,,
\label{ximoments}
\end{align}
which are linear combinations of Gegenbauer coefficients of order $\leq
n$. This corresponds to an expansion of the LCDA in terms of a
delta-function and its derivatives,   
\begin{align}
 \phi_\pi(u;\mu) & {} = 2 \, \sum_{n=0}^\infty \, \frac{(-1)^n}{n!} \,
 \delta^{(n)}(2u-1) \;\, \langle \xi^n  \rangle_\pi (\mu)  \,.
\end{align}
Results for the first few moments $\langle \xi^n \rangle_\pi$ are shown
in Table~\ref{tab:ximom} for the strict non-relativistic limit,
including the NLO corrections from (\ref{eq:NRLCDABNLO}) and comparing
with the non-relativistic corrections of order $v^2_{\rm NR}$ discussed
by Braguta et al.~in \cite{Braguta:2006wr}. Keeping first-order
corrections in $v_{\rm NR}^2$ only, this formally amounts to the
replacement    
\beq
  \phi_\pi^{\rm NR}(u) &{}=& \delta(u-1/2) + \frac{v_{\rm NR}^2}{24} \,
  \delta''(u-1/2)   + {\cal O}(v_{\rm NR}^4) \,.
\eeq
In particular, this fixes the moment
$ \langle \xi^2 \rangle_\pi = \frac{v_{\rm NR}^2}{3} $.
The authors \cite{Braguta:2006wr} propose a resummed formula, 
\beq
  \phi_\pi^{\rm NR}(u) &\to& 
\frac{1}{v_{\rm NR}} \, \theta\left(u-\frac{1-v_{\rm NR}}{2}\right) \, 
                        \theta\left(\frac{1+ v_{\rm NR}}{2}-u\right) \,.
\label{phiNRexp2}
\eeq
The comparison in Table~\ref{tab:ximom} shows that for $v_{\rm NR}^2
\simeq \alpha_s(m)\simeq 0.2$, the effect of the $v^2_{\rm NR}$
corrections is qualitatively and quantitatively similar to the
$\alpha_s$ corrections from (\ref{eq:NRLCDABNLO}).

\begin{table}[t!!bp]
\begin{center}
\parbox{0.9\textwidth}{\caption{\small The moments $\langle \xi^n
    \rangle_\pi$ at the non-relativistic scale
    $\mu=m$. \label{tab:ximom}}  
\vspace{1em}}
\begin{tabular}{|l | c c c c c|}
\hline
 {$n$} & { 2}  & { 4}  & { 6}  & { 8} & {10}
\\
\hline \hline
 NR limit & 0 & 0 & 0 & 0 &0 \\
 NLO (\ref{eq:NRLCDABNLO}) (for $\alpha_s=0.2$) 
 % & 3.139 & 0.496 & 0.176 & 0.085 & 0.048
 & 0.067 & 0.011 & 0.004 & 0.002 & 0.001
\\
 $v^2_{\rm NR}$ (\ref{phiNRexp2}) (for $v^2_{\rm NR}=0.2$)
& 0.067 & 0.008 & 0.001 & 0.000 & 0.000
\\
\hline
\end{tabular}
\end{center}
\end{table}

It is also interesting to determine the correction to the first inverse
moment of the LCDA which appears in QCD factorization formulas
\begin{align}
\langle u^{-1}\rangle_\pi^{(1)}(\mu) & {}\equiv
\int_0^1 \! du \;\; \frac{\phi_\pi^{(1)}(u;\mu)}{u} 
\simeq  3 \left( 2.73 +1.08 \, \ln\frac{\mu^2}{m^2} \right) \,.
\end{align}
Finally, we quote the result for the derivative of $\phi_\pi(u)$ at the
endpoints 
\begin{align}
  \phi'_\pi(0;\mu)=-\phi'_\pi(1;\mu)&{}= \frac{\alpha_s C_F}{4\pi} 
 \left( 4 + 12 \ln \frac{\mu^2}{m^2} \right) + {\cal O}(\alpha_s^2),
\end{align}
which is sometimes discussed in the context of non-factorizable
contributions to hard exclusive reactions
\cite{Bagan:1997bp,Arnesen:2006vb}.  

For non-equal quark masses, the NLO corrections to the $\overline{\rm
  MS}$\/-renormalized twist-2 LCDA are given by 
\begin{align}
\phi_K^{(1)}(u;\mu) &{}= 2  \left\{\!\!
\left( \ln \frac{\mu^2}{m_K^2 \, (u_0-u)^2} -1 \right)\!\!
 \left[ \left( 1+\frac{1}{u_0-u} \right) \frac{u}{u_0} \;\theta(u_0-u) 
 + \left( 
   \begin{array}{c} 
     u \leftrightarrow \bar u \\
     u_0 \leftrightarrow \bar u_0 
   \end{array}
 \right)
 \right] \right\}_{+} \no \\
& \quad 
+4  \left\{ \frac{u (1-u)}{(u_0-u)^2} 
\right\}_{++}
+ 2 \, \delta'(u-u_0) \left( 2 u_0 (1-u_0) \,\ln \frac{u_0}{1-u_0} +2
  u_0 -1 \right). 
\end{align}
The first moment now becomes
\begin{align}
\int_0^1 \! du \; u\,\phi_K(u;\mu) &{}= u_0 + \frac{\as C_F}{4\pi} 
\left[ \left( -\frac43 \, \ln \frac{\mu^2}{u_0^2 \, m_K^2}
-\frac{7 (1-u_0)}{3} \, \ln u_0^2 -\frac{38}{9} \right) u_0 
- ( u_0 \leftrightarrow \bar u_0 ) \right]\,.
\label{eq:phi:first}
\end{align}

\subsubsection{2-particle LCDAs of twist-3}

The twist-3 LCDAs for the 2-particle Fock states are obtained in the
same way as the twist-2 one. After absorbing the corrections to the
local matrix elements into the renormalized values for $\mu_\pi$ and
$\tilde \mu_\pi$, we obtain a UV-divergent piece
\begin{align}
\phi_p^{(1)}(u) \big|_{\rm div.} &{}=   
\frac{2}{\vareps} \left[ \left( 1+\frac{1}{u_0-u} \right)
  \;\theta(u_0-u)  
 + \left( 
  \begin{array}{c} 
     u \leftrightarrow \bar u \\
     u_0 \leftrightarrow \bar u_0 
   \end{array}
 \right) \right]_{+}
\end{align}
and a finite NLO contribution to the twist-3 LCDA associated to the
pseudoscalar current 
\begin{align}
\phi_p^{(1)}(u;\mu) &{}= 2  \left\{\!\! 
\left(\ln \frac{\mu^2}{m_K^2(u_0-u)^2} -1
\right)\!\!
  \left[ \left( 1+\frac{1}{u_0-u} \right) \;\theta(u_0-u) 
 + \left( 
  \begin{array}{c} 
     u \leftrightarrow \bar u \\
     u_0 \leftrightarrow \bar u_0 
   \end{array}
 \right) \right] \right\}_{+} \no \\
& \quad 
+4 u_0 (1-u_0) \left( \left\{ \frac{1}{(u_0-u)^2} 
 \right\}_{++} + \delta'(u-u_0)  \,\ln \frac{u_0}{1-u_0} \right)
+2 \left\{ \frac{2 u_0-1}{(u_0-u)} \right\}_{+} \,.
\end{align}
In particular, the first moment of $\phi_p(u)$ now reads
\begin{align}
\int_0^1 \! du \; u\,\phi_p(u;\mu) &{}= 
u_0 +\frac{\as C_F}{4\pi} 
\left[ \left( -3 \, \ln \frac{\mu^2}{m_K^2} 
+6 u_0 \ln u_0 -4 \right) u_0 
- ( u_0 \leftrightarrow \bar u_0 ) \right],
\label{eq:phiP:first}
\end{align}
which is in agreement with the eom-constraint from
(\ref{eq:eom:local2}). At the endpoints we now have
\begin{align}
 \phi_p(0;\mu) & {} = \frac{\alpha_s C_F}{4\pi } \left(
  \frac{2+2u_0}{u_0} \, \ln \frac{\mu^2}{m_1^2} - 2\right)
 + {\cal O}(\alpha_s^2) 
\end{align}
and similar for $\phi_p(1;\mu)$ with $m_1 \leftrightarrow m_2$, i.e.\
$u_0 \leftrightarrow \bar{u}_0$. For the twist-3 LCDA associated to the
pseudo\-tensor current (whose normalization factor starts at order
$\alpha_s$), we simply have
\begin{align}
\phi_\sigma(u) &{}= 2 \left[ \frac{u}{u_0} \,\theta(u_0-u)  
+ \left( 
   \begin{array}{c} 
     u \leftrightarrow \bar u \\
     u_0 \leftrightarrow \bar{u}_0 
   \end{array}
 \right) \right] + \calO(\as).
\end{align}
In contrast to the other 2-particle LCDAs in (\ref{delta}), we find that
$\phi_\sigma(u)$ is not given by a delta-like distribution in the NR
limit and has support for $0\leq u\leq1$.

\subsection{Heavy mesons}

The calculation of the LCDAs for a $B_c$ meson (which again can be
considered as a toy model for LCDAs of $B_q$ mesons with $m_b \gg m_q$)
goes along the same lines as for the $\eta_c$ case. However, important
differences arise because the heavy $b$-quark is to be treated in HQET
which modifies the divergence structure of the loop integrals (notice
that in our set-up, a charm quark in a $B_c$ meson is treated as
"light"). As a consequence, the evolution equations for the LCDAs of
heavy mesons \cite{Lange:2003ff} differ from those of light mesons.

\subsubsection{The LCDA $\mathbf \phi_B^+(\omega)$}

Let us first focus on the distribution amplitude $\phi_B^+(\om)$ which
enters the QCD factorization formulas for exclusive heavy-to-light
decays. In the local limit we derive the corrections from soft gluon
exchange to the decay constant in HQET. We find
\begin{align}
\hat{f}_M(\mu) &{} = f_M^\text{NR} \left[ 1 + \, \frac{\alpha_s
    C_F}{4\pi} 
\left( 3\ln\frac{\mu}{m}-4 \right)+ {\cal O}(\alpha_s^2) \right].
\end{align}
Notice that the decay constant of a heavy meson exhibits the well-known
scale dependence in HQET \cite{Neubert:1993mb}. The remaining NLO
corrections to the distribution amplitude $\phi_B^+(\omega)$ contain an
UV-divergent piece (details of the derivation can be found in
Appendix~\ref{app:phim})   
\begin{align} \label{phiplus:UV}
 \phi^{(+,1)}_B(\om;\mu) \big|_{\rm div.} &= 
\frac{2\om}{\epsilon} \left[  
\frac{\theta(m-\om)}{m(m-\om)}+\frac{\theta(\om-m)}{\om(\om-m)}
  \right]_+ 
- \delta(\om-m) \left[\frac{1}{\eps^2} - \frac{1}{\eps}
  \left( 1- \ln \frac{\mu^2}{m^2}\right) \right] 
\end{align}
and a finite piece
\begin{align} 
\frac{\phi^{(+,1)}_B(\om;\mu)}{\om} 
& {}= 
2  \left[ \left(\ln \left[\frac{\mu^2}{(\om-m)^2}\right]-1\right) \left(
    \frac{\theta(m-\om)}{m (m-\om)} 
+ \frac{\theta(\om-m)}{\om(\om-m)}\right)
\right]_+ 
+ 
4  \left[ \frac{\theta(2m-\om)}{(\omega-m)^2} \right]_{++} 
\no \\
& {}
+ \frac{4 \, \theta(\om-2m)}{(\om-m)^2} 
- \frac{\delta(\om-m)}{m} \left( \frac12 \ln^2
\frac{\mu^2}{m^2} -  \ln \frac{\mu^2}{m^2}  +
\frac{3\pi^2}{4} + 2 \right) 
\label{phiplus}
\end{align}
with an analogous definition of plus-distributions as
in~(\ref{eq:plusdef},\ref{eq:plusplusdef}). Notice that, in order to
separate the UV divergence coming from the longitudinal momentum
integration, we have introduced an  auxiliary parameter $\mu_f \equiv
2m$ to split the support region of the LCDA into two parts. The
distribution amplitude in (\ref{phiplus}) obeys the evolution equation  
\begin{align} \label{eq:NRLCDABEvol}
\frac{d}{d \ln \mu} \; \phi_B^+(\om;\mu) & {} =  - \, \frac{\alpha_s
  C_F}{4\pi} \int_0^\infty d\om' \; \gamma_+^{(1)}(\om,\om';\mu) \;
\phi_B^+(\om';\mu)  + {\cal O}(\alpha_s^2) \,,
\end{align}
where the anomalous dimension $\gamma_+^{(1)}(\om,\om';\mu)$ can be read
off the UV-divergent terms in (\ref{phiplus:UV}) and is given by
\cite{Lange:2003ff}  
\begin{align}
\gamma_+^{(1)}(\om,\om';\mu) & {}=
\left( \Gamma_{\rm cusp}^{(1)} \, \ln \frac{\mu}{\om} - 2
\right) \delta(\om-\om') - \Gamma_{\rm cusp}^{(1)} \, \om \left[
\frac{\theta(\om'-\om)}{\om'(\om'-\om)} +
\frac{\theta(\om-\om')}{\om(\om-\om')} \right]_+
\label{LN-kernel}
\end{align}
with $\Gamma_{\rm cusp}^{(1)}=4$.

In contrast to the light meson case, the normalization of the heavy
meson distribution amplitude is ill-defined. Imposing a hard cutoff
$\Lambda_{\rm UV}\gg m$ and expanding to first order in $m/\Lambda_{\rm
  UV}$, we find
\begin{align}
\int\limits_0^{\Lambda_{\rm UV}} d\om \; \phi_B^+(\om;\mu)
& {} \simeq 1 - \frac{\alpha_s C_F}{4\pi} \left[ \frac12 \ln^2
\frac{\mu^2}{\Lambda_{\rm UV}^2}+\ln \frac{\mu^2}{\Lambda_{\rm UV}^2} +
\frac{\pi^2}{12} \right] + {\cal O}(\alpha_s^2) + {\cal
O}(m/\Lambda_{\rm UV})
\end{align}
and similarly for the first moment
\begin{align}
\frac{1}{\Lambda_{\rm UV}} \,
\int\limits_0^{\Lambda_{\rm UV}} d\om \; \om\, \phi_B^+(\om;\mu)
& {} \simeq \frac{\alpha_s C_F}{4\pi} \left[ 2\ln
  \frac{\mu^2}{\Lambda_{\rm UV}^2} +6 \right] + {\cal O}(\alpha_s^2)+
{\cal O}(m/\Lambda_{\rm UV}) \,. 
\end{align}
The last two expressions provide model-independent properties of the
distribution amplitude which have been studied within the operator
product expansion in~\cite{Lee:2005gz}. Our results are in agreement
with these general findings.  

We finally quote our result for two phenomenologically relevant moments
in the factorization approach to heavy-to-light decays
\cite{Braun:2003wx,Lee:2005gz} 
\begin{align} 
(\lambda_B(\mu))^{-1} \equiv \int\limits_0^\infty d\om \;\;
\frac{\phi_B^+(\om;\mu)}{\om} & {} = \frac{1}{m} \left(
1 -\frac{\alpha_s C_F}{4\pi} \left[ \frac12
\ln^2 \frac{\mu^2}{m^2} - \ln \frac{\mu^2}{m^2} +\frac{3\pi^2}{4} -2
\right] \right) +{\cal O}(\alpha_s^2)\,,
\label{lamBdef}
\end{align}
and
\begin{align}
\sigma_B(\mu) & {} \equiv \sigma_B^{(1)}(\mu)
 = \ln \frac{\mu}{m} 
+ \frac{\alpha_s C_F}{4\pi} \left[8 \zeta_3\right]
+ {\cal O}(\alpha_s^2) \,,
\label{sigBNR}
\end{align}
where $\zeta_j = \sum_{n=1}^\infty n^{-j}$ is the Riemann zeta function
and we defined  
\begin{align}\label{sigBdef}
\sigma_B^{(n)}(\mu)
 &{} \equiv \lambda_B(\mu) \, \int\limits_0^\infty d\om \,
\frac{\phi_B^{+}(\om;\mu)}{\om} \left[ \ln \frac{\mu}{\omega} \right]^n
\,. 
\end{align}
The leading-order scale-dependence of these quantities is in general
given by  
\begin{align}
 \frac{d \lambda_B^{-1}}{d\ln\mu} & {} = - \frac{\alpha_s C_F}{4 \pi} 
 \left(\Gamma_{\rm cusp}^{(1)} \, \sigma_B - 2 \right) \, \lambda_B^{-1}
 +{\cal O}(\alpha_s^2)\,,
\\
\frac{d \sigma_B}{d\ln\mu} & {} = 1 + \frac{\alpha_s C_F}{4 \pi}\,
  \Gamma_{\rm cusp}^{(1)} \left( (\sigma_B)^2 - \sigma_B^{(2)} \right) 
  +{\cal O}(\alpha_s^2)\,.
\end{align}
In particular, in the non-relativistic limit $\sigma_B^{(n)}(\mu) =
\left( \sigma_B(\mu) \right)^n$ and therefore the $\alpha_s$ correction
on the right hand side of (\ref{sigBNR}) does not depend explicitly on
$\ln \mu$. For arbitrary values of $n$, we find for the scale dependence
of the logarithmic moments
\begin{align}
\frac{d \sigma_B^{(n)}}{d \ln \mu}  &{}=  
n \sigma_B^{(n-1)} + \frac{\as C_F}{4\pi} \; \Gamma_\text{cusp}^{(1)} 
\left[ \sigma_B^{(1)} \, \sigma_B^{(n)} - \sigma_B^{(n+1)} + 2n!
  \sum_{j=1}^{[n/2]} \frac{\zeta_{2j+1}}{(n-2j)!} \, \sigma_B^{(n-2j)}
\right] + \calO(\as^2),
\label{eq:Bplus:momentsren}
\end{align}
where $[x]$ denotes the greatest integer less than or equal to $x$.

\subsubsection{The LCDA $\mathbf \phi_B^-(\omega)$}

A similar analysis can be performed for the other 2-particle LCDA of the
$B$ meson. We now obtain for the UV-divergent piece (see also
Appendix~\ref{app:phim}) 
\begin{align}
\phi^{(-,1)}_B(\om;\mu)\big|_{\rm div.} & {} = 
\frac{2}{\epsilon} \left[ 
  \frac{\theta(m-\om)}{(m-\om)} \right]_+ 
+ \frac{2}{\eps} \, \frac{\theta(m-\om)}{m}
+ \frac{2\om}{\epsilon}  \left[   \frac{\theta(\om-m)}{\om(\om-m)}
\right]_+ 
\CR & \quad
  - \delta(\om-m) \left[\frac{1}{\eps^2} + \frac{1}{\eps}
 \left( 1+ \ln \frac{\mu^2}{m^2} \right) 
\right] \,.
\label{phiminus:UV}
\end{align}
The finite contributions read
\begin{align} 
\phi^{(-,1)}_B(\om;\mu) 
&{}= 
2 \left[ \left(\ln \left[\frac{\mu^2}{(\om-m)^2}\right]-1\right) 
\frac{\theta(m-\om)}{m-\om} \right]_+
+ 2  \left(\ln \left[\frac{\mu^2}{(\om-m)^2}\right]-1\right) 
\frac{\theta(m-\om)}{m}
\nonumber \\[0.2em]
&{} + 2 \om \left[ \left(\ln
    \left[\frac{\mu^2}{(\om-m)^2}\right]-1\right)
  \frac{\theta(\om-m)}{\om(\om-m)} \right]_+ 
+ 
4 m \left[ \frac{\theta(2m-\om)}{(\omega-m)^2} \right]_{++} 
\no \\
&{}
+ 4m \,\frac{\theta(\om-2m)}{(\om-m)^2} 
- \delta(\om-m) \left( \frac12 \ln^2
\frac{\mu^2}{m^2} +  \ln \frac{\mu^2}{m^2}  +
\frac{3\pi^2}{4} + 6 \right).
\label{phiminus}
\end{align}
The distribution amplitude in (\ref{phiminus}) obeys the evolution
equation  
\begin{align} \label{eq:minusvol}
\frac{d}{d \ln \mu} \; \phi_B^-(\om;\mu) = {} &   - \, \frac{\alpha_s
  C_F}{4\pi} \int_0^\infty d\om' \; \gamma_-^{(1)}(\om,\om';\mu) \;
\phi_B^-(\om';\mu) 
\cr
 &  -\, \frac{\alpha_s C_F}{4\pi}
\int_0^\infty d\om' \; \gamma_{-+}^{(1)}(\om,\om';\mu) \;
\phi_B^+(\om';\mu) + {\cal O}(\alpha_s^2) \,,
\end{align}
where the anomalous dimension kernels $\gamma_-^{(1)}(\om,\om';\mu)$ and
$\gamma_{-+}^{(1)}(\om,\om';\mu)$ can be read off the UV-divergent terms
in (\ref{phiminus:UV}) (see Appendix~\ref{app:phim} for details) 
\begin{align} \label{gammaminuskernel}
\gamma_-^{(1)}(\om,\om';\mu) & {}=
\gamma_+^{(1)}(\om,\om';\mu) - \Gamma_{\rm cusp}^{(1)} \,
\frac{\theta(\om'-\om)}{\om'} \,, 
\\[0.2em]
\gamma_{-+}^{(1)}  (\om,\om';\mu) & {}=
- \Gamma_{\rm cusp}^{(1)} \,\left[\frac{m \, \theta
    (\om'-\om)}{\om'{}^2}\right]_+  \,.
\end{align}
Among others, the knowledge of $\gamma_-$ is essential to check the
factorization of certain correlation functions appearing in sum-rule
calculations for $B \to \pi$ form factors within SCET
\cite{DeFazio:2005dx}.  

Another new result are the first positive moments of the LCDA
$\phi_B^-(\omega)$ as a function of a hard cutoff $\Lambda_{\rm UV}\gg
m$,   
\begin{align}
\int\limits_0^{\Lambda_{\rm UV}} d\om \; \phi_B^-(\om;\mu)
&{}\simeq 1 - \frac{\alpha_s C_F}{4\pi} \left[ \frac12 \ln^2
\frac{\mu^2}{\Lambda_{\rm UV}^2} - \ln \frac{\mu^2}{\Lambda_{\rm UV}^2}
+ \frac{\pi^2}{12} \right] + {\cal O}(\alpha_s^2) + {\cal
O}(m/\Lambda_{\rm UV})\,, 
\label{phimmom1}
\\[0.3em]
\frac{1}{\Lambda_{\rm UV}} \,
\int\limits_0^{\Lambda_{\rm UV}} d\om \; \om\, \phi_B^-(\om;\mu)
&{}\simeq \frac{\alpha_s C_F}{4\pi} \left[ 2\ln
  \frac{\mu^2}{\Lambda_{\rm UV}^2} +2 \right] + {\cal O}(\alpha_s^2)+
{\cal O}(m/\Lambda_{\rm UV}) \,, 
\label{phimmom2}
\end{align}
which are again expected to be model-independent. Actually, these
moments can already be derived in the Wandzura-Wilczek
approximation. From the solution of (\ref{ww0}) in $D=4-2\epsilon$
dimensions  
\begin{align}
\phi_B^-(\om;\mu) &=
(1-\epsilon)
\int_\om^\infty d\eta\; \frac{\eta-m}{\eta^{2}} \,
\left(\frac{\eta}{\om}\right)^\epsilon \; \phi_B^+(\eta;\mu)
+ \frac{m}{\om} \phi_B^+(\om;\mu),
\end{align}
we obtain the bare (unrenormalized) moments
\begin{align}
 \langle \omega^n \rangle_B^-  \; \simeq \; \frac{\Lambda^{n+1} \,
   \phi_B^-(\Lambda) + (1-\epsilon) \; \langle \omega^n
   \rangle_B^+}{n+1-\epsilon},
\label{moments:WW}
\end{align}
which result in the same $\overline{\rm MS}$\/-subtracted moments as in
(\ref{phimmom1},\ref{phimmom2}), i.e.~the 3-particle LCDAs only
contribute subleading terms to these moments in our case.  

We finally quote the quantity
\begin{align}
 \phi_B^-(0;\mu)&{} = \frac{\alpha_s C_F}{4\pi} \, \frac{4}{m} \, \ln
 \frac{\mu^2}{m^2} + {\cal O}(\alpha_s^2)\,,
\end{align}
which plays a role in sum-rule calculations for heavy-to-light form
factors \cite{DeFazio:2005dx,Khodjamirian:2005ea}.

\subsubsection{3-particle LCDAs and equations of motion}

In order to verify whether the equations of motion
(\ref{eom1},\ref{eom2}) hold after including first order relativistic
corrections, we have to compute the 3-particle LCDAs which arise at
order $\alpha_s$ in the non-relativistic limit. Without going into
details, we quote our results for the bare LCDAs that enter
(\ref{eom1},\ref{eom2})   
\begin{eqnarray}
\Psi_V(\omega,\xi) &=& \frac{\as C_F}{4\pi}
\frac{\delta(\om-m+\xi)}{2m} 
\bigg\{\left( \frac{1}{\vareps} + \ln \frac{\mu^2}{\xi^2} -1 \right)
\xi^2 \,\theta(m-\xi) - \left( \frac{1}{\vareps} + \ln \frac{\mu^2}{m^2}
  +1 \right) m^3\,\delta(\xi-m)  \bigg\},\nonumber\\[0.5em]
\Psi_A(\omega,\xi) &=& \frac{\as C_F}{4\pi} 
\frac{\delta(\om-m+\xi)}{2m} 
\bigg\{\!\!-\!\left( \frac{1}{\vareps} + \ln \frac{\mu^2}{\xi^2} +1 \right)
\xi^2 \,\theta(m-\xi) -\! \left( \frac{1}{\vareps} + \ln \frac{\mu^2}{m^2}
  +1 \right) m^3\,\delta(\xi-m)  \bigg\},\nonumber\\[0.5em] 
X_A(\omega,\xi) &=&  \frac{\as C_F}{4\pi} \;
\bigg[ 2 \left( \frac{1}{\vareps} + \ln \frac{\mu^2}{\xi^2} \right)
\xi\,\delta(\om-m) - \frac{\delta(\om-m+\xi) }{2m}  \nonumber\\[0.3em]
& & \;
\bigg\{ \left[ (4m-3\xi) \left( \frac{1}{\vareps} + \ln
    \frac{\mu^2}{\xi^2} \right) -\xi \right] \xi\, \theta(m-\xi) - \left(
  \frac{1}{\vareps} + \ln \frac{\mu^2}{m^2} +1 \right) m^3\delta(\xi-m)
\bigg\} \bigg].
\label{3particle}
\end{eqnarray}
We show in Appendix~\ref{app:eom} that the eom-constraint (\ref{eom1})
is indeed fulfilled to order $\alpha_s$. On the other hand we find that
(\ref{eom2}) \emph{does not} hold beyond tree level since the
$\xi$--integral involving our result for the 3-particle LCDA $X_A$ is
ill-defined for $\xi\to\infty$. Since we again expect the radiative tail
of the 3-particle LCDAs (which determines the large-$\xi$ behaviour) to be
model-independent, the failure of (\ref{eom2}) beyond tree level should
be considered a general feature.

\section{Renormalization-group evolution}

\label{sec5}

In physical applications, the light-cone distribution amplitudes are
required at the hard-scattering scale $\mu$ which is set by the momentum
transfer in the exclusive reaction. The evolution from the ``soft''
scale $m$ to the hard-scattering scale $\mu$ resums large logarithms
$\ln \mu^2/m^2$. In this section we study the evolution of the NR
distribution amplitudes to leading logarithmic (LL) approximation. For
''light'' mesons we focus on the twist-2 LCDA $\phi_\pi(u)$ and for
''heavy'' mesons we consider the 2-particle LCDAs $\phi_B^+(\omega)$ and
$\phi_B^-(\omega)$.

\subsection{The twist-2 LCDA $\phi_\pi(u)$}

The evolution of the twist-2 LCDA $\phi_\pi(u;\mu)$ is described by the
Brodsky-Lepage kernel (\ref{ERBL}). To solve the evolution equation
(\ref{EE:phipi}), one projects the distribution amplitude onto
Gegenbauer polynomials which are the eigenfunctions of the evolution
kernel,       
\beq
  \phi_\pi(u;\mu) &{}=& 6 u \bar{u} \left( 1 + \sum_{n=1}^\infty
       a_{n}(\mu) \; C_n^{3/2}(2u-1) \right) \,.
\label{conformal}
\eeq
The respective coefficients are obtained from (\ref{proj}) and have the
LL evolution    
\beq
  a_{n}(\mu) &{}=& a_n(\mu_0)
    \left(\frac{\alpha_s(\mu)}{\alpha_s(\mu_0)}\right)^{-\gamma_{n}/\beta_0}
\,, \quad \gamma_{n} = C_F \left(
     3 + \frac{2}{(n + 1)(n + 2)} - 4 \sum_{j=1}^{n+1}\frac{1}{j} \right),
\label{evol}
\eeq
with $\beta_0 = (33-2n_f)/3$ (for illustration, we will use $n_f=3$ in
the numerical examples). For very large $n \gg 1$ we have 
$
  \gamma_n  \simeq  - 4 \, C_F  \ln n \,,
$
which implies that the effect of higher Gegenbauer coefficients becomes
less important at high scales. 

\begin{table}[tpbt]
\begin{center}
\parbox{0.9\textwidth}{\caption{\small 
 The moments $\langle \xi^n \rangle_\pi$ as a function of the evolution
 parameter $\eta=\alpha_s(\mu)/\alpha_s(m)$. \label{tab:ximom2}}    
\vspace{1em}}
\begin{tabular}{|l | c c c c c|}
\hline
 {$n$} & {2}  & {4}  & {6}  & {8} & {10}
\\
\hline \hline
 LL ($\eta=1/5$) 
 &0.126 & 0.048 & 0.025 & 0.015 & 0.010
\\
 LL ($\eta=1/25$) 
 & 0.173 & 0.070 & 0.038 & 0.024 &0.016
\\
asymptotic 
& 0.200 & 0.086 & 0.048 & 0.030 & 0.021
\\ \hline
\end{tabular}
\end{center}
\end{table}
 
We show in Table~\ref{tab:ximom2} the LL evolution of the moments
$\langle \xi^n \rangle_\pi$ defined in (\ref{ximoments}), starting from
the tree-level result in the NR limit, $\phi_\pi(u;\mu_0=m) =
\delta(u-1/2)$, for two values of $\eta=\alpha_s(\mu)/\alpha_s(m)$ and
in the asymptotic limit. In contrast to the moments $\langle \xi^n
\rangle_\pi$, the phenomenologically important $1/u$ moment is a linear
combination of an \emph{infinite} number of Gegenbauer moments,   
\beq
\langle u^{-1} \rangle_\pi(\mu) = 3 \sum_{j=0}^\infty a_{2j}(\mu).
\eeq
In order to study the evolution effects from the non-relativistic scale,
where $\langle u^{-1} \rangle_\pi(\mu_0=m) = 2$, towards $\mu \gg m$, it
will therefore be crucial to control the effects of higher Gegenbauer
coefficients. For this purpose, we find it convenient to consider model
parameterizations which are obtained from a slight modification of the
strategy developed in~\cite{Ball:2005ei}. Our ansatz involves three real
parameters $a > 0$, $b>0$, $0 \leq t_c \leq 1$,   
\beq
  \phi_\pi(u) & \equiv &  \frac{3 u \bar u}{\Gamma(a;-\ln t_c)}
  \, \int_0^{t_c} dt \, (- \ln t)^{a-1} 
  \left[ f(2u-1,i t^{1/b})+ f(2u-1,-i t^{1/b}) \right] 
\label{model}
\eeq
with $\Gamma(a;b)= \int_b^\infty dt \; t^{a-1} e^{-t}$ and the
generating function of the Gegenbauer polynomials, 
\beq
  f(\xi,\theta) &{}=& \frac{1}{(1 - 2 \xi \theta + \theta^2)^{3/2}} 
  \;=\;
 \sum_{n=0}^\infty C_n^{3/2}(\xi) \, \theta^n \,.
\eeq
Performing the $t$\/-integration in (\ref{model}), one finds
\beq
  \phi_\pi(u) & = & 6u\bar{u} \;\;\sum_{n=0}^\infty
  \left[ \cos \left(\frac{n \pi}{2}\right) \, 
        \frac{\Gamma(a;\, -(1+n/b) \, \ln t_c)}{\Gamma(a;-\, \ln t_c)}
\left(n/b +1 \right)^{-a} \right] \, 
C_n^{3/2}(2u-1) \,,
\eeq
from which one reads off the Gegenbauer coefficients $a_{\rm odd}=0$ and
\beq
  a_n &{}=& \frac{(-1)^{n/2}}{(n/b+1)^a} \;
  \frac{\Gamma(a;\, -(1+n/b) \, \ln t_c)}{\Gamma(a;-\, \ln t_c)}
  \qquad {\rm for}~n~{\rm even}.
\eeq
For $t_c \to 0$ our ansatz reduces to the asymptotic distribution
amplitude and for $t_c \to 1$ it is equivalent to one of the models
discussed in~\cite{Ball:2005ei}, where the Gegenbauer coefficients show a
simple power-like fall-off (in this case with alternating signs). As
observed in~\cite{Ball:2005ei}, for values of $a\leq 3$, the model
induces some pathological behaviour at $u=1/2$. In our ansatz this is
regularized by the cut-off parameter $t_c < 1$. The qualitative
behaviour of the Gegenbauer coefficients for large $n$ now depends on
$t_c$:  
\begin{itemize}
 \item For moderately large values of $n$, we have
  \beq
  1 \ll n \ll n_{\rm crit} \equiv - b \left(1+1/\ln t_c\right) \quad &:& \quad
   |a_n| \simeq (n/b)^{-a}\,,
  \eeq
  i.e.\ a power-like fall-off with $n$ as in~\cite{Ball:2005ei}.

\item For asymptotically large values of $n$, one obtains
\beq
  n \gg n_{\rm crit} &:& \quad
   |a_n| \simeq \frac{b \,(-\ln t_c)^{a-1}}{\Gamma(a;-\ln t_c)} \;
          \frac{t_c^{n/b}}{n} \,,
  \eeq
  i.e.\ an exponential fall-off with $n$, which renders the
  contribution of {\em very} high Gegenbauer coefficients irrelevant.
\end{itemize}
We now reconsider the evolution of the tree-level result in the NR limit
$\phi_\pi(u;\mu_0=m) = \delta(u-1/2)$ and fix the model parameters
$(a,b,t_c)$ in (\ref{model}) from the first three non-vanishing
Gegenbauer coefficients using (\ref{proj}),     
\vspace*{-3mm}
\beq\left.
\begin{array}{l}
 a_2(m) \,\simeq\, -0.5833 \\
 a_4(m) \,\simeq\, +0.4583 \\
 a_6(m) \,\simeq\, -0.3906
\end{array} \quad \right\} &\quad \Rightarrow\quad  & 
\begin{array}{l}
 a \,\simeq\, 0.3962 \\
 b \,\simeq\, 0.8045 \\
 t_c \,\simeq\, 0.9993\\[0.2em]
 n_{\rm crit} \,\simeq\, 1149
\end{array}
\eeq
The fact that $a < 1$ and $n_{\rm crit} \gg 1$ reflects the bad
convergence of the Gegenbauer expansion in the NR limit. Still, the
model parameterization reproduces the Gegenbauer coefficients with
$n\ll n_{\rm crit}$ and the value of the first inverse moment $\langle
u^{-1}\rangle_\pi$ to a very good accuracy (see the first line
in Table~{\ref{tab:evol}}).

\begin{figure}[t!bph]
\linespread{1.1}
\begin{center}
\centerline{ \psfrag{phipi}{
\hspace{-1em}$\phi_\pi(u;\mu)$}
 \psfrag{u}{$u$}
 \epsfig{file=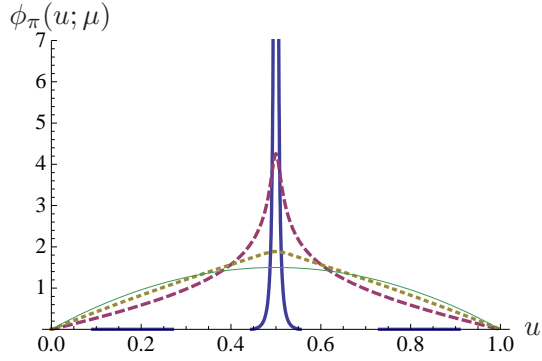, width=0.4\textwidth}}
\vspace{2mm}
\centerline{\parbox{0.9\textwidth}{\caption{\label{fig:phiM} {\small
        Approximation of the NR distribution amplitude
        $\phi_\pi(u)=\delta(u-1/2)$ in terms of the model
        parameterization (\ref{model}) (thick solid line) and its
        evolution for $\eta = 1/5$ (dashed line) and $\eta=1/25$ (dotted
        line). The asymptotic LCDA is shown for comparison (thin solid
        line).}}}} 
\end{center}
\vspace{-2mm}
\end{figure}

The same strategy can be applied for scales $\mu>m$. The LL evolution
towards larger scales depends on  
$
  \eta_i = \alpha_s(\mu_i)/\alpha_s(m).
$
For illustration, we consider $\eta_1=1/5$ and $\eta_2=1/25$ and obtain
\beq\left.
\begin{array}{l}
 a_2(\mu_1) \,\simeq\, -0.2160 \\
 a_4(\mu_1) \,\simeq\, +0.1079 \\
 a_6(\mu_1) \,\simeq\, -0.0679
\end{array} \quad \right\} &\quad \Rightarrow\quad  & 
\begin{array}{l}
 a \,\simeq\, 1.2679 \\
 b \,\simeq\, 0.8708 \\
 t_c \,\simeq\, 0.9811 \\[0.2em]
 n_{\rm crit.}\,\simeq\, 45
\end{array}
\eeq
and
\vspace*{-2mm}
\beq\left.
\begin{array}{l}
 a_2(\mu_2) \,\simeq\, -0.0800 \\
 a_4(\mu_2) \,\simeq\, +0.0254 \\
 a_6(\mu_2) \,\simeq\, -0.0118
\end{array} \quad \right\} &\quad \Rightarrow\quad  & 
\begin{array}{l}
 a \,\simeq\, 2.1451 \\
 b \,\simeq\, 0.8966 \\
 t_c \,\simeq\, 0.9418 \\[0.2em]
 n_{\rm crit.}\,\simeq\, 14
\end{array}
\eeq
We observe that the parameter $a$ increases under evolution, which is
related to the growth of the anomalous dimensions for larger values of
$n$, leading to a steeper fall-off of the Gegenbauer coefficients at
larger scales. Effectively, for moderately large values of $n$, one has
\beq
  a(\mu_i) \approx a(m) - \frac{4 C_F}{\beta_0} \, \ln \eta_i \,.
\eeq
The parameter $b$ is only slightly increasing while $t_c$ is decreasing
under evolution. The critical value of $n$ is quickly decreasing from
$n_{\rm crit}\simeq1149$ at $\mu=m$ to $n_{\rm crit}\simeq14$ at
$\mu=\mu_2$. Figure~\ref{fig:phiM} shows the evolution of the model LCDA
as a function of $u$. For $\eta=1/5$ the functional form still
``remembers'' the non-relativistic profile, while for $\eta=1/25$ it is
already close to the asymptotic form. Table~\ref{tab:evol} compares the
first few Gegenbauer coefficients using the exact projection of the
delta-function and the model parameterization (\ref{model}). We see that
the differences are tiny and the model gives a good approximation. We
also quote the result for the first inverse moment which slowly evolves
from the NR value, $\langle u^{-1}\rangle_\pi(\mu_0=m)=2$, towards the
asymptotic value $\langle u^{-1}\rangle_\pi(\mu\to\infty)=3$. We clearly
see that the model gives a better description for relatively low scales
than a truncated conformal expansion (\ref{conformal}) with $n\leq6$
(i.e.~with the same number of input parameters as our model). The latter
can be improved, however, by considering the averaged moment  
\beq
\langle u^{-1} \rangle_\pi(\mu) \simeq
\frac32 \left( \sum_{j=0}^{j_{max}} a_{2j}(\mu) +
\sum_{j=0}^{j_{max}-1} a_{2j}(\mu) \right),
\eeq
which accounts for the alternating sign behaviour of the Gegenbauer
coefficients. Using this improved truncated conformal expansion, we find
that the $1/u$ moment is given by
$
\langle u^{-1}\rangle_\pi=(2.04,2.57,2.82)
$
at respective scales $(m,\mu_1,\mu_2)$, which is very similar to the
predictions of the model parameterization (see Table~{\ref{tab:evol}}).

\begin{table}[t]
\linespread{1.1}
\begin{center}
\parbox{0.9\textwidth}{\caption{\label{tab:evol} \small LL evolution of
    the first few Gegenbauer coefficients starting from the NR
    distribution amplitude $\phi_\pi(u)=\delta(u-1/2)$. For each value
    of the evolution parameter $\eta=\alpha_s(\mu)/\alpha_s(m)$ we show
    the results for the exact expression, the model parameterization
    (\ref{model}) and a truncated conformal expansion (\ref{conformal})
    with $n\leq6$. We also quote the value for the first inverse moment
    $\langle u^{-1}\rangle_\pi$.}       
\vspace{1em}}
\scriptsize\tiny\footnotesize\small
\begin{tabular}{| l || c | c | c | c | c | c | c | c | c || c |}
\hline
 & $a_2$    & $a_4$    & $a_6$    & $a_8$    & $a_{10}$ 
 & $a_{12}$ & $a_{14}$ & $a_{16}$ & $\langle u^{-1}\rangle_\pi$
\cr
\hline\hline
exact ($\eta=1$)  & 
-0.583 & 0.458 & -0.391 & 0.346 & -0.314 & 
0.290 & -0.271 & 0.255 & 
\cr
model & 
$*$ &  $*$ & $*$ &  0.346 & -0.314 & 
 0.289   & -0.269 & 0.253   & 2.00
\cr
conformal  exp.& $*$ &  $*$ & $*$ &
\multicolumn{5}{c|}{(truncation $n\leq 6$)} 
& 1.45
\cr
\hline\hline
exact($\eta=1/5$)  & 
-0.216 & 0.108 & -0.068 & 0.048 & -0.036 & 
0.029 & -0.023 & 0.020 & 
\cr
model & 
$*$ &  $*$ & $*$ &  0.048 & -0.036 & 
 0.028   & -0.023 & 0.019  & 2.55 
\cr
conformal  exp.& $*$ &  $*$ & $*$ &
\multicolumn{5}{c|}{(truncation $n\leq 6$)} 
& 2.47 
\cr
\hline\hline
exact($\eta=1/25$)  & 
-0.080 & 0.025 & -0.012 & 0.007 & -0.004 & 
0.003 & -0.002& 0.002 & 
\cr
model & 
$*$ &  $*$ & $*$ &  0.007 & -0.004 & 
 0.003   & -0.002 & 0.001 & 2.81 
\cr
conformal  exp.& $*$ &  $*$ & $*$ &
\multicolumn{5}{c|}{(truncation $n\leq 6$)} 
& 2.80
\cr\hline
\end{tabular}
\end{center}

\end{table}

\subsection{The LCDA $\phi_B^+(\omega)$}

\begin{figure}[t!pbt]
\linespread{1.1}
\begin{center}
\psfrag{phiplus}{$m \, \phi_B^+(\omega;\mu)$}
\psfrag{om}{$\om/m$}
 \includegraphics[width=0.38\textwidth]{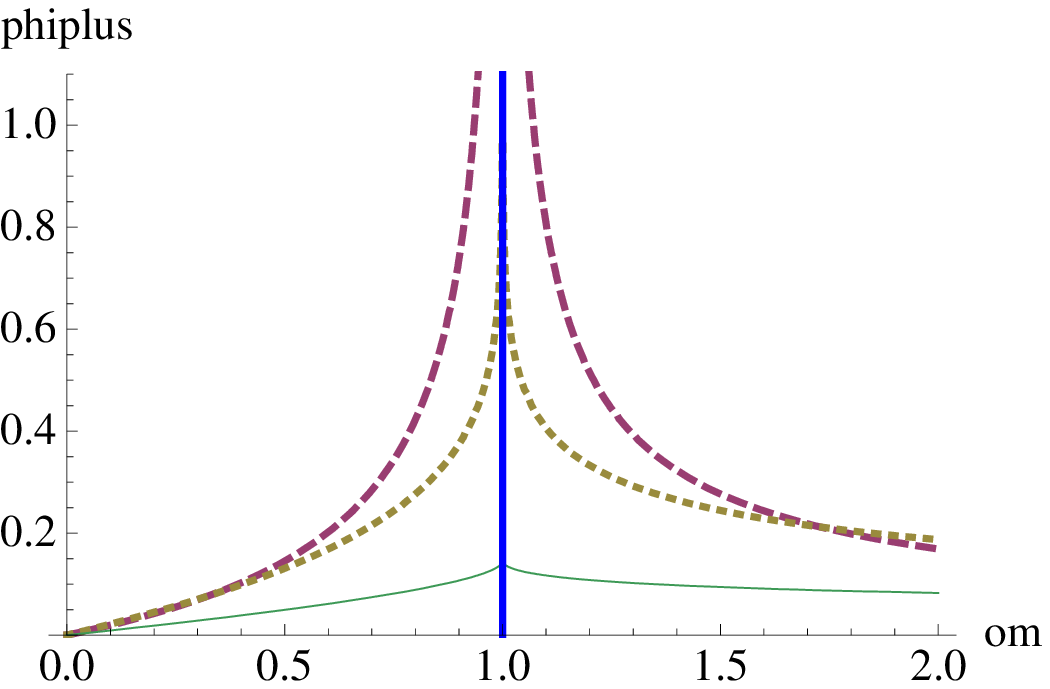} \ \ \ \ \
 \includegraphics[width=0.405\textwidth]{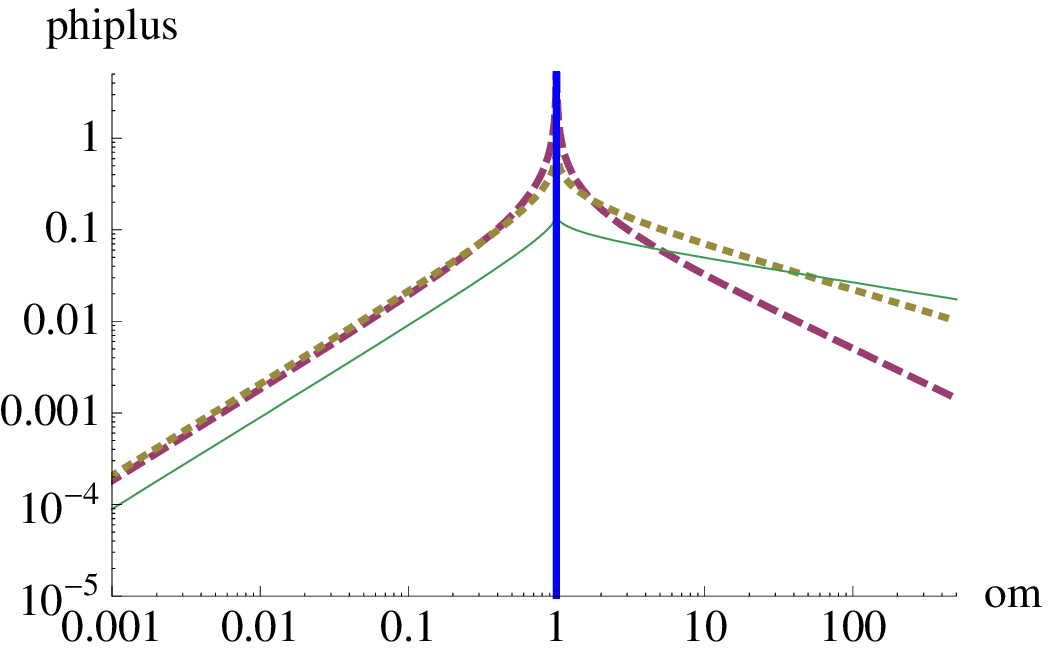} 
\parbox{0.9\textwidth}{\caption{\label{fig:phipevol} \small Evolution of
    the heavy meson LCDA $\phi_B^+(\omega;\mu)$ starting from the
    tree-level result $\phi_B^+(\omega;\mu_0=m)=\delta(\omega-m)$, where
    we assumed $\alpha_s(m)=1$ (thick solid line). The curves (dashed,
    dotted, thin solid line) correspond to
    $\eta=\alpha_s(\mu)/\alpha_s(m)=1/2,1/5,1/10$, respectively.}}  
\\[1.5em]
\psfrag{phiplus}{$m \, \phi_B^+(\omega;\mu)$}
\psfrag{om}{$\om/m$}
 \includegraphics[width=0.38\textwidth]{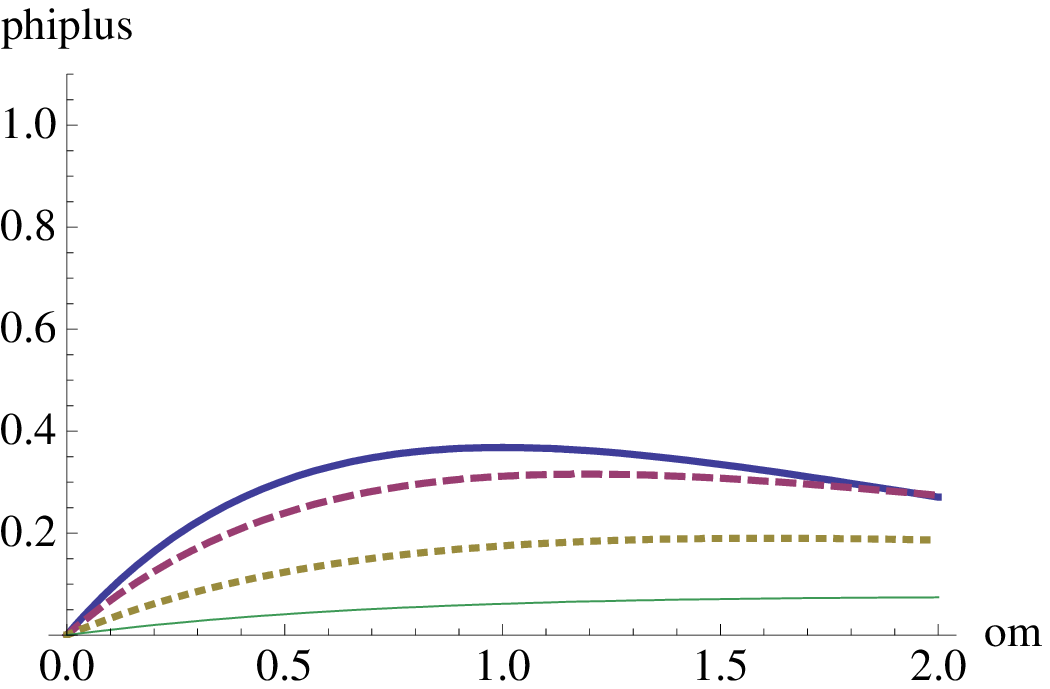} \ \ \ \ \
 \includegraphics[width=0.405\textwidth]{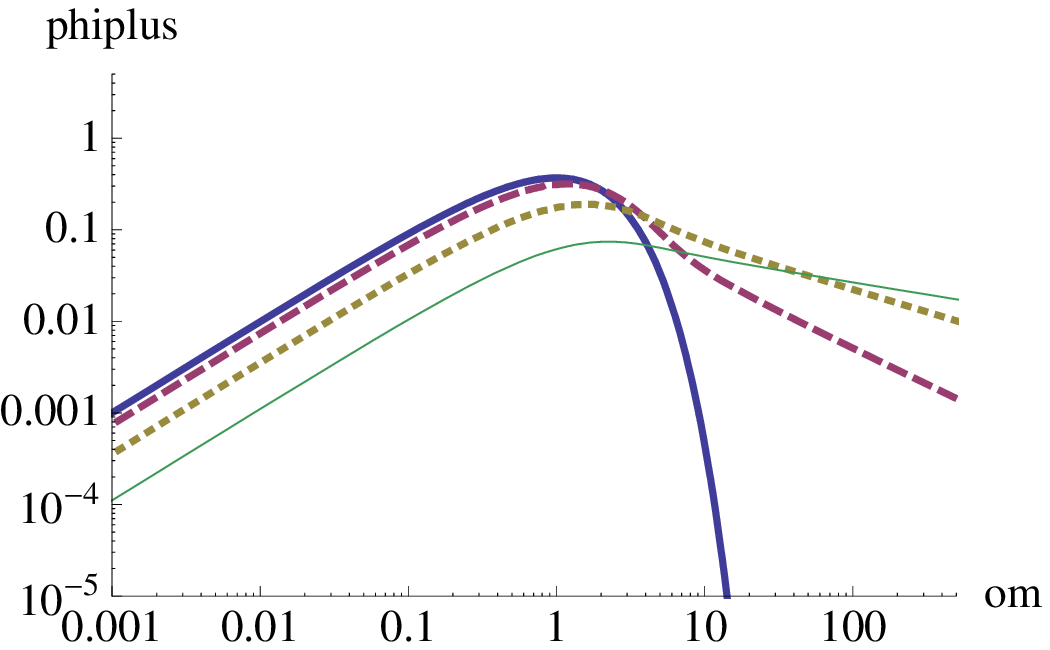} 
\vspace{0.8em}
\parbox{0.9\textwidth}{\caption{\label{fig:phipevol_exp} \small The same
    as Figure~\ref{fig:phipevol} with initial condition
    $\phi_B^+(\omega;\mu_0)= \omega/m^2 \,e^{-\omega/m}\,$(thick solid
    line).}}      
\end{center}
%\vspace{-5mm}
\end{figure}

The evolution of the LCDA $\phi_B^+(\omega;\mu)$ for scales $m \leq \mu
\leq M$ is described by the Lange-Neubert kernel
(\ref{LN-kernel}).\footnote{Notice that above the $b$\/-quark mass scale
  one should match the LCDAs in HQET onto LCDAs in QCD, see
  e.g.~\cite{Pilipp:2007sb}.} The solution of the evolution equation
(\ref{eq:NRLCDABEvol}) can be written in closed form
as~\cite{Lee:2005gz}    
\begin{eqnarray}
   \phi_B^+(\omega;\mu)
   &{}=& e^{V -2 \, \gamma_E \, g}\,\frac{\Gamma(2-g)}{\Gamma(g)}
    \int_0^\infty\!\frac{d\omega'}{\omega'}\,\phi_B^+(\omega';\mu_0)
 \left( \frac{\omega_>}{\mu_0} \right)^g \frac{\omega_<}{\omega_>}  
    \ {}_2F_1\Big(1-g,2-g; 2; \frac{\omega_<}{\omega_>}\Big) \,,
\label{phipevol}
\end{eqnarray}
where $\omega_<=\min(\omega,\omega')$ and
$\omega_>=\max(\omega,\omega')$. The evolution is controlled by the
functions    
\begin{equation}
   V\equiv V(\mu,\mu_0)
   = - \int\limits_{\alpha_s(\mu_0)}^{\alpha_s(\mu)}\!\!
   \frac{d\alpha}{\beta(\alpha)} \Bigg[ \Gamma_{\rm cusp}(\alpha)
   \!\!\int\limits_{\alpha_s(\mu_0)}^\alpha\!\!
   \frac{d\alpha'}{\beta(\alpha')} + \gamma(\alpha) \Bigg] \,,
\label{Vdef}
\end{equation}
with $\Gamma_{\rm cusp} \simeq \frac{\alpha_s C_F}{\pi}$, $\gamma \simeq
- \frac{\alpha_s C_F}{2\pi}$,
$\beta\simeq-\frac{\alpha_s^2\beta_0}{2\pi}$ (we use $n_f=4$ in the
numerical examples) and    
\begin{equation}
   g\equiv g(\mu,\mu_0)
   = \int\limits_{\alpha_s(\mu_0)}^{\alpha_s(\mu)}\!\!\!d\alpha\,
    \frac{\Gamma_{\rm cusp}(\alpha)}{\beta(\alpha)}
   \simeq \frac{2C_F}{\beta_0} \ln\frac{\alpha_s(\mu_0)}{\alpha_s(\mu)} \,.
\label{gdef}
\end{equation}
The hypergeometric function $\,{}_2F_1(a,b;c;z)$ has the series
expansion 
$$
\,{}_2F_1(a,b;c;z) = \sum_{n=0}^\infty \,
\frac{\Gamma(a+n) \Gamma(b+n) \Gamma(c)}{\Gamma(a) \Gamma(b) \Gamma(c+n)}
\, \frac{z^n}{n!} \,.
$$
Starting from the tree-level result in the non-relativistic limit
$\phi_B^+(\omega;\mu_0=m) = \delta(\omega-m)$, we obtain for scales $\mu
> m$ the relatively simple expression 
\begin{eqnarray}\label{tree}
   m \, \phi_B^+(\omega;\mu) \Big|_{\rm tree}
   &{}=& 
     e^{V -2 \, \gamma_E \, g}\,\frac{\Gamma(2-g)}{\Gamma(g)}
   \left( \frac{\omega_>}{m} \right)^g \frac{\omega_<}{\omega_>}  
    \ {}_2F_1\Big(1-g,2-g; 2; \frac{\omega_<}{\omega_>}\Big) \,,
\label{phipevolsol}
\end{eqnarray}
where now $\omega_<=\min(\omega,m)$, $\omega_>=\max(\omega,m)$,
$g=g(\mu,m)$ and $V=V(\mu,m)$. Fixing the value of $\alpha_s(m)$ at the
NR input scale, we may study how the shape of $\phi_B^+(\omega;\mu)$ is
changed by evolution effects. In Figure~\ref{fig:phipevol} we have
plotted (\ref{phipevolsol}) for $\alpha_s(m)=1$ and three different
values of $\eta=\alpha_s(\mu)/\alpha_s(m)$. As expected, the evolution
drives the initial delta-function shape towards a flatter
distribution. In the double-logarithmic plot on the right-hand side in
Figure~\ref{fig:phipevol}, we may read off the asymptotic behaviour of
$\phi_B^+(\omega;\mu)$ for $\omega \to 0$ and $\omega\to\infty$. As
argued on general grounds~\cite{Lange:2003ff}, the LCDA develops a
linear behaviour for $\omega \to 0$, whereas for $\omega \to \infty$ it
tends to fall off slower than $1/\omega$ at higher scales. This can also
be seen by comparison with Figure~\ref{fig:phipevol_exp}, where we plot
the evolution of another LCDA with initial condition
$\phi_B^+(\om;\mu_0=m) = \om/m^2 e^{-\om/m}$. 

\begin{figure}[t!pbt]
\linespread{1.1}
\begin{center}
\psfrag{lambdaBdm}{$\lambda_B(\mu)/\lambda_B(\mu_0)$}
\psfrag{sigmaBt}{$\sigma_B(\mu)-\ln \frac{\mu}{\mu_0}$}
\psfrag{eta}{$\ \eta(\mu)$}
 \includegraphics[width=0.4\textwidth]{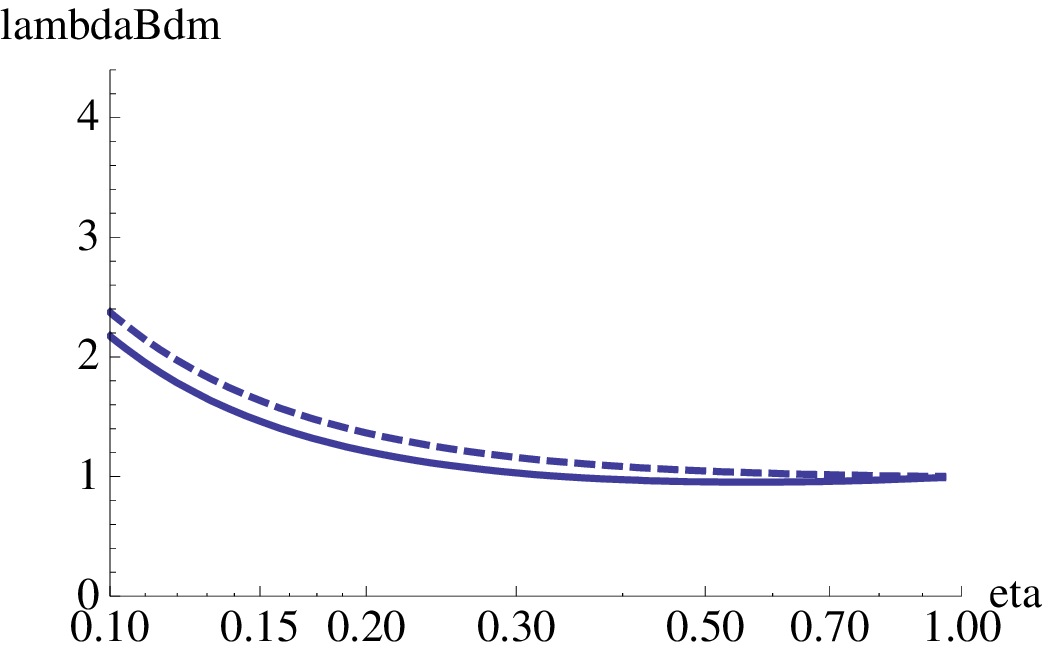} 
\ \ \ \ \ \ \
 \includegraphics[width=0.4\textwidth]{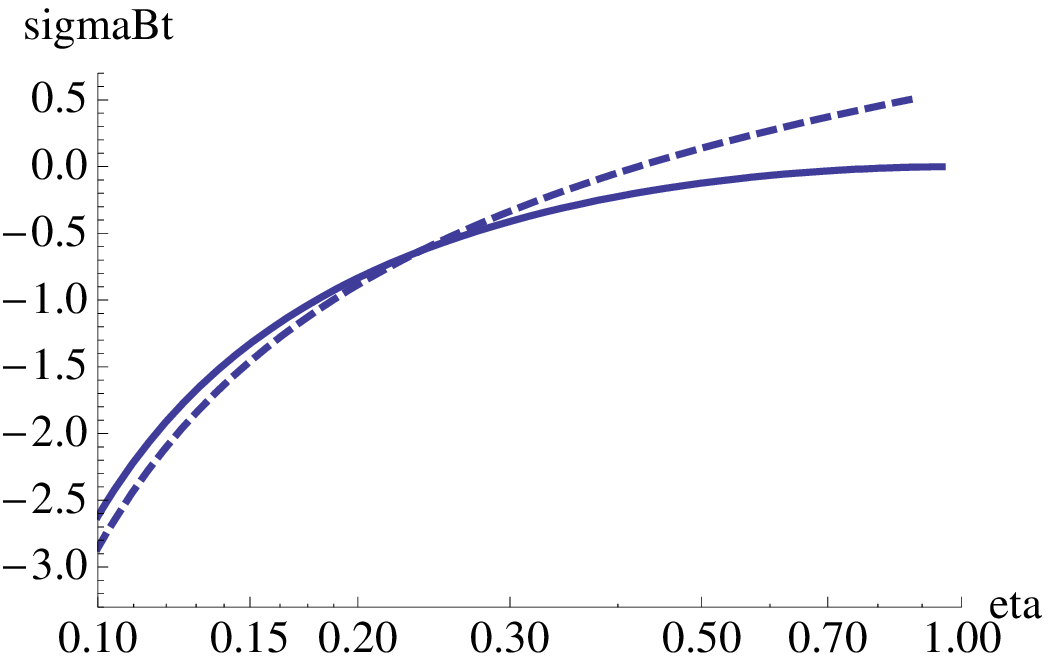} 
\vspace{0.8em}
\parbox{0.9\textwidth}{\caption{\small The moments
    $\lambda_B(\mu)/\lambda_B(\mu_0)$ and $\sigma_B(\mu)-\ln\mu/\mu_0$
    as a function of the evolution parameter
    $\eta=\alpha_s(\mu)/\alpha_s(\mu_0)$. The solid line refers to the
    initial condition $\phi_B^+(\om;\mu_0=m)=\delta(\om-m)$ and the
    dashed line to $\phi_B^+(\om;\mu_0=m)= \omega/m^2 \,e^{-\omega/m}\,$,
    where we assumed $\alpha_s(m)=1$. 
\label{fig:lamBevol}}}
\end{center}
\end{figure}

In Figure~\ref{fig:lamBevol} we show the corresponding evolution of the phenomenologically relevant moments $\lambda_B(\mu)$ and $\sigma_B(\mu)$ defined in (\ref{lamBdef},\ref{sigBdef}).
From (\ref{phipevol}) we find the closed formulas
\begin{align}
\frac{1}{\lambda_B(\mu)} &=
 e^{V-2 \,\gamma_E g}  \, 
\frac{\Gamma(1-g)}{\Gamma(1+g)}
 \, \int_0^\infty \frac{d\omega}{\om} \left(\frac{\om}{\mu_0}\right)^g 
    \phi_B^+(\om;\mu_0) \,, \\[0.5em]
\sigma_B(\mu) &=
   g \, (1-g) \, \,_4F_3(1,1,1-g,2-g;2,2,2;1) \no\\[0.15em]
& 
 \quad {} - \frac{g}{1-g} \, \,_3F_2(1-g,1-g,1-g;2,2-g;1) \no \\[0.05em]
& \quad {} -
\left(\int_0^\infty \frac{d\om}{\om} \left(\frac{\om}{\mu_0}\right)^g  \phi_B^+(\om;\mu_0) \right)^{-1}
 \int_0^\infty \frac{d\om}{\om} \left(\frac{\om}{\mu_0}\right)^g
 \ln \frac{\om}{\mu} \ \phi_B^+(\om;\mu_0) \,.
\end{align}
In general, the evolution of the moments $\lambda_B$ and $\sigma_B$ thus depends
on the shape of the LCDA $\phi_B^+(\omega)$ \cite{Lange:2003ff}. 
For our examples, $\phi_B^+(\om;\mu_0)=\delta(\om-\mu_0)$,
respectively $\phi_B^+(\om;\mu_0)=\om/\mu_0^2 \, e^{-\om/\mu_0}$,
the $\omega$ integration can be performed explicitly,
leading to relatively simple analytic expressions. It is also possible
to approximate the factors 
$(\om/\mu_0)^g = 1 + g \, \ln(\om/\mu_0) + \ldots$
In this approximation, the evolution for the moment $\lambda_B(\mu)$ can be entirely determined in 
terms of $\lambda_B(\mu_0)$ and $\sigma_B(\mu_0)$, see also \cite{Braun:2003wx}.

% \begin{align}
% \frac{\lambda_B(\mu)}{\lambda_B(\mu_0)} &=
%  e^{2 \,\gamma_E g -V}  \, 
% \frac{\Gamma(1+g)}{\Gamma(1-g)} \,, \\[0.3em]
% \sigma_B(\mu)-\ln \frac{\mu}{\mu_0} &=
%    g \, (1-g) \, \,_4F_3(1,1,1-g,2-g;2,2,2;1) \no\\
% & 
%  \quad {} - \frac{g}{1-g} \, \,_3F_2(1-g,1-g,1-g;2,2-g;1) \,.
% \end{align}
% For the initial condition $\phi_B^+(\om;\mu_0=m)=\om/m^2 e^{-\om/m}$ we obtain similar analytical results,
% \begin{align}
% \frac{\lambda_B(\mu)}{\lambda_B(\mu_0)} &=
%  e^{2 \,\gamma_E g -V} \, \frac{1}{\Gamma(1-g)} \,, 
% \\[0.3em]
% \sigma_B(\mu)-\ln \frac{\mu}{\mu_0} &=
%    g \, (1-g) \, \,_4F_3(1,1,1-g,2-g;2,2,2;1) \no\\
% & 
% \quad {} - \frac{g}{1-g} \, \,_3F_2(1-g,1-g,1-g;2,2-g;1)
% %\no\\ & \quad
%  - \psi(1+g) \, .
% \end{align}

\subsection{The LCDA $\phi_B^-(\omega)$}

\begin{figure}[t!pbtp]
\linespread{1.1}
\begin{center}
\psfrag{phiminus}{$m \, \phi_B^-(\omega,\mu)$}
\psfrag{om}{$\om/m$}
\includegraphics[width=0.38\textwidth]{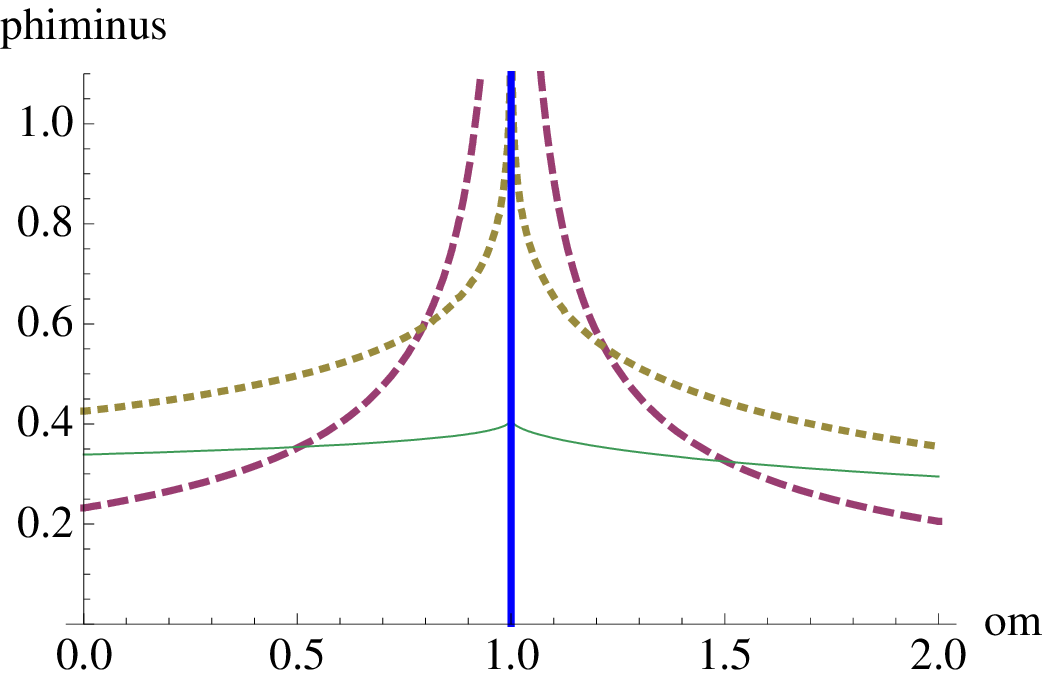} \ \ \ \ \
 \includegraphics[width=0.405\textwidth]{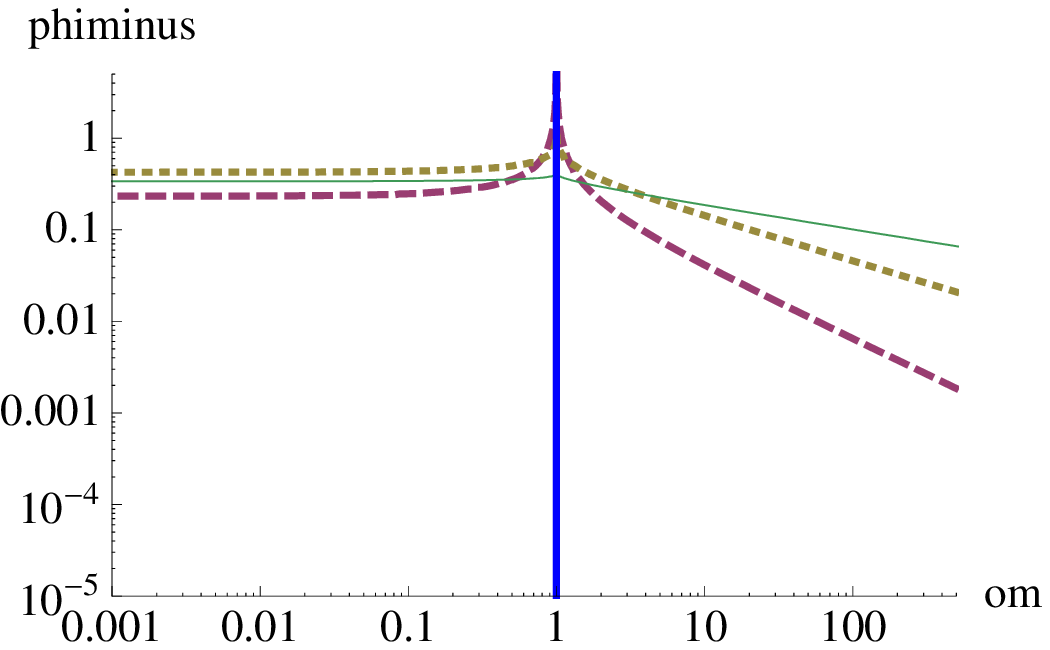} 
\\[1em]
\includegraphics[width=0.38\textwidth]{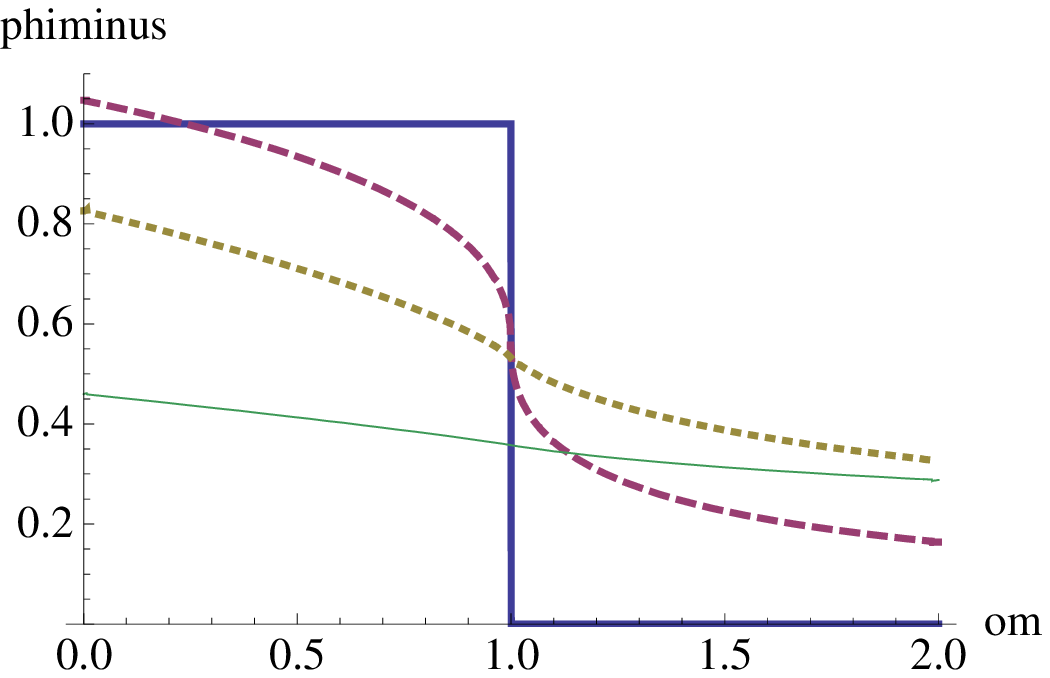} \ \ \ \ \
 \includegraphics[width=0.405\textwidth]{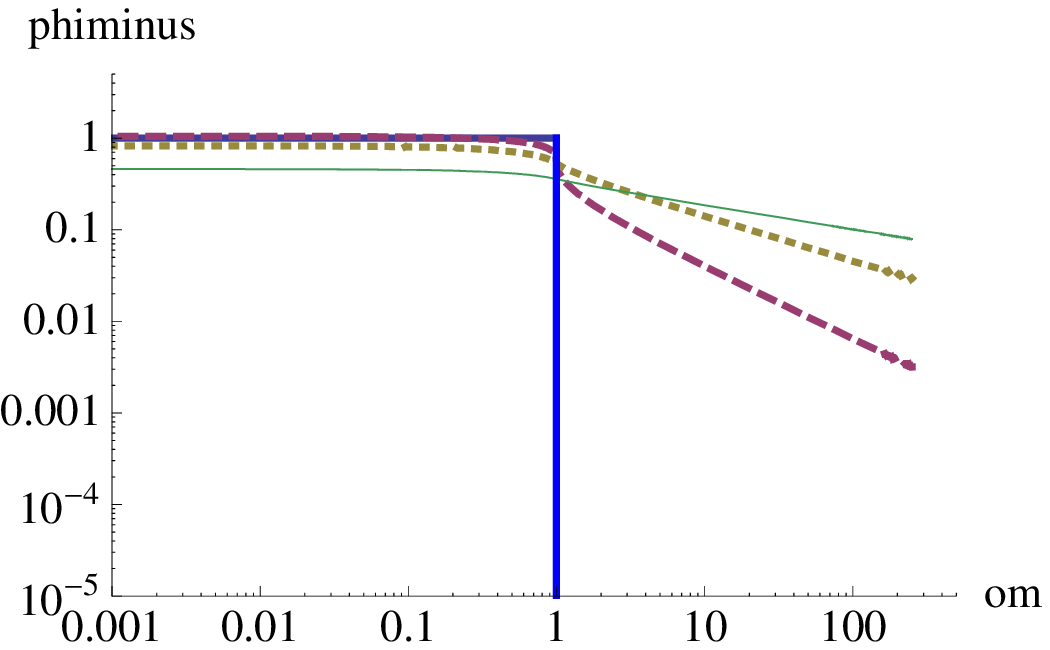} 
\\[1em]
 \includegraphics[width=0.38\textwidth]{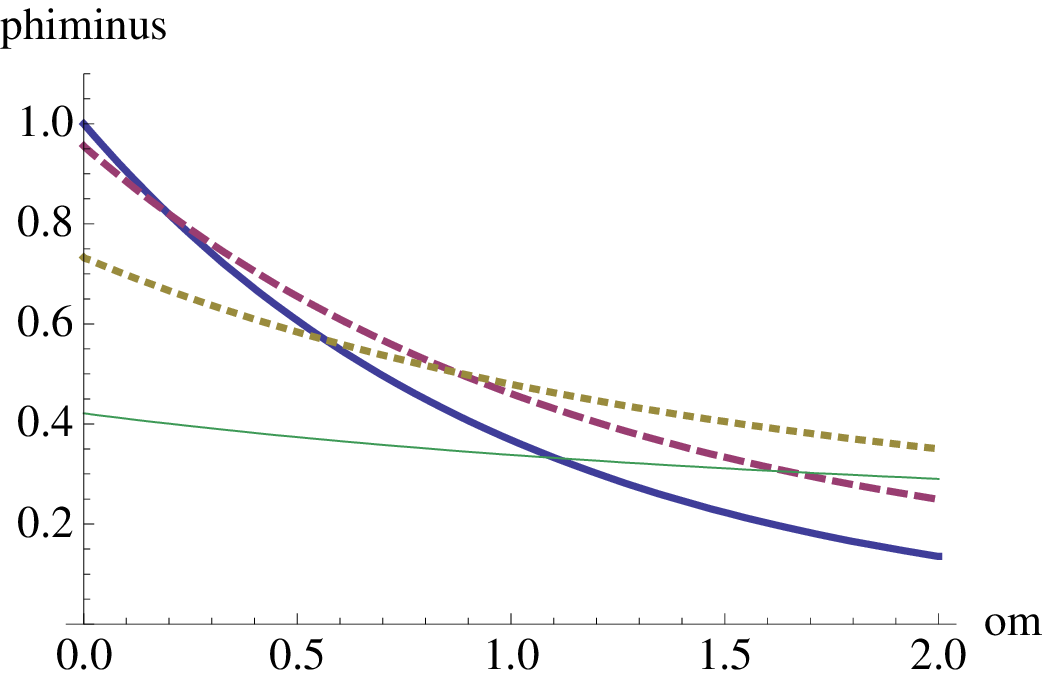} \ \  \ \ \
 \includegraphics[width=0.405\textwidth]{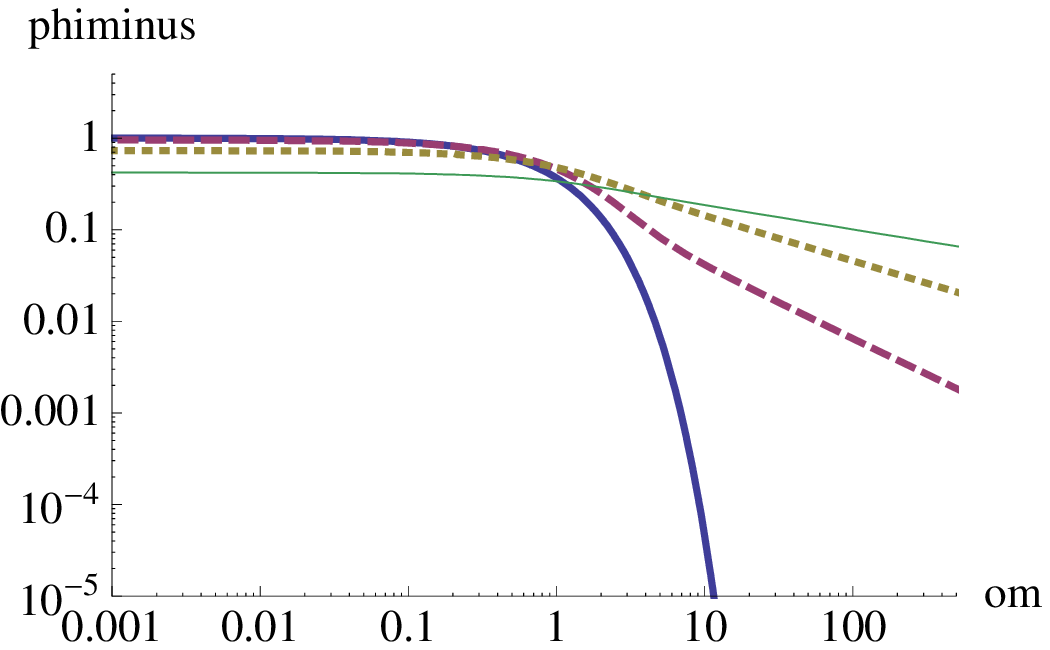} 
\vspace{0.8em}
\parbox{0.9\textwidth}{\caption{\label{fig:phimevol} \small Evolution of
    the LCDA $\phi_B^-(\om;\mu)$. Notations and conventions as in
    Figures~\ref{fig:phipevol},\ref{fig:phipevol_exp}. The initial
    conditions at the scale $\mu_0=m$ are $\phi_B^-(\omega;\mu_0)=
    \delta(\om-m)$ (upper row), $\phi_B^-(\omega;\mu_0)=
    \theta(m-\om)/m$ (middle row) and
    $\phi_B^-(\om;\mu_0)=1/m\,e^{-\om/m}$ (lower row).}} 
\end{center}
\end{figure}

The evolution of the LCDA $\phi_B^-(\om;\mu)$ is somewhat more involved,
because of the possible mixing with the 3-particle LCDAs. In addition,
for a non-vanishing light quark mass $m \neq 0$, we have seen in
(\ref{eq:minusvol}) that the LCDA $\phi_B^+(\om;\mu)$ mixes into 
$\phi_B^-(\om;\mu)$. In the following, we concentrate on possible
applications in realistic $B_q$ decays (where we can set $m=0$),
neglecting the contributions from 3-particle LCDAs which is left for
future work. In this approximation, the solution of the evolution
equation 
\begin{align} 
\frac{d}{d \ln \mu} \; \phi_B^-(\om;\mu) & {} =  - \, \frac{\alpha_s
  C_F}{4\pi} \int_0^\infty d\om' \; \gamma_-^{(1)}(\om,\om';\mu) \;
\phi_B^-(\om';\mu)  + {\cal O}(\alpha_s^2) 
\nonumber \\[0.2em]
 & \quad \ {} + \left\{
 \begin{array}{l} \mbox{\footnotesize terms with $m\neq 0$} \\
    \mbox{\footnotesize contributions from 3-particle LCDAs}
 \end{array} \right.
\label{evolapprox}
\end{align}
can be obtained in a similar way as for
$\phi_B^+(\om;\mu)$~\cite{Lange:2003ff,Lee:2005gz}. The details of the
derivation can be found in Appendix~\ref{app:evol}. As a result, the
solution for $\phi_B^-(\om;\mu)$ can be written as  
\begin{align}
 \phi_B^-(\om;\mu)  \;{} \simeq \;{} &
e^{V - 2 \, \gamma_E \, g} \,
\frac{\Gamma(1-g)}{\Gamma(g)} 
\int_0^\infty
\frac{d\om'}{\om_>} \ \phi_B^-(\om';\mu_0) \,
\left(\frac{\om_>}{\mu_0}\right)^g \ 
  {}_2F_1\Big(1-g,1-g,1,\frac{\om_<}{\om_>}\Big) \,.
\label{solapprox}
\end{align}
In Figure~\ref{fig:phimevol} we illustrate the evolution of
$\phi_B^-(\omega;\mu)$ for three different initial conditions at the
scale $\mu_0=m$:  
\begin{itemize}
 \item $\phi_B^-(\om;\mu_0) = \delta(\om-m)$,
 \item $\phi_B^-(\om;\mu_0) = \theta(m-\om)/m$,
 \item $\phi_B^-(\om;\mu_0) = 1/m\,e^{-\om/m}$.
\end{itemize}
The first example corresponds to the strict non-relativistic limit
(where the neglect of the light quark mass in the evolution equation may
be considered as inconsistent). The second and third example follow from
the Wandzura-Wilczek relation (\ref{ww0}) for the initial LCDAs
$\phi_B^+(\om;\mu_0)$ considered in the previous subsection. While the
behaviour of $\phi_B^-(\om;\mu)$ at small values of $\omega$ depends on
the model for the initial distribution, the radiative tail for large
values of $\om$ is again universal. More precisely, the solution
(\ref{solapprox}) of the (approximate) evolution equation suggests that
$\phi_B^-(\om;\mu)$ also falls off slower than $1/\om$ at higher scales,
while the slope of the LCDA at $\om=0$ tends to vanish under evolution,
independent of the initial behaviour of the distribution amplitude.

%\clearpage

\section{Summary}

Non-relativistic $q\bar q$ bound states have been used as a starting
point to construct light-cone distribution amplitudes for light mesons
in QCD and heavy mesons in HQET. At the non-relativistic scale, the
leading 2-particle distribution amplitudes can be approximated by delta
functions, fixing the light-cone momenta of the quarks according to
their masses. After including radiative corrections from relativistic
gluon exchange, the distribution amplitudes cover the whole physically
allowed support region, $0\leq u\leq1$ for light mesons and $0 \leq
\omega <\infty$ for heavy mesons. In this paper, explicit expressions
for 2-particle distribution amplitudes of twist-2 and twist-3 for
''light'' mesons (with quark masses $m_1\sim m_2$) have been calculated
to first order in the strong coupling constant. In the same way,
next-to-leading order expressions for the 2- and 3-particle distribution
amplitudes have been derived for ''heavy'' mesons (where $m_1\gg
m_2$). We also studied the evolution of the 2-particle distribution
amplitudes under change of renormalization scale.  

Our results apply to the physical situation of a hard exclusive reaction,
that involves bound states of heavy bottom or charm quarks, with large
momentum transfer, for instance, 
$B_c \to \eta_c \ell\nu$
\cite{Bell:2006tz,Huang:2007kb,Hernandez:2006gt,Ivanov:2006ni,Ebert:2003cn}, 
$e^+e^- \to J/\psi \, \eta_c$  
\cite{Choi:2007ze,Bodwin:2006dm,Bondar:2004sv,Ma:2004qf}
or $\gamma^*\gamma \to \eta_c$ 
\cite{Feldmann:1997te}. 
Moreover, from
the divergence structure of our explicit next-to-leading order results,
we could derive certain model-independent properties which also hold for
bound states of relativistic quarks. In this way we used our calculation
as a toy model to derive new results for the $B$ meson distribution
amplitude $\phi_B^-$, as the cut-off dependence of positive moments, the
anomalous dimension kernel and the solution of the evolution equation in
the Wandzura-Wilczek approximation. The toy model also allowed us to
address an issue that has been controversial in the literature, i.e.~the
question if the constraints from the equations of motion hold beyond
tree level in the heavy meson case.

\section*{Acknowledgements}

T.F.\ is supported by the German Ministry of Research (BMBF, contract
No.~05HT6PSA). He also acknowledges financial support by the 
Cluster of Excellence ``Origin and Structure of
the Universe'' during his stay at the TU Munich in fall 2007. The work
of G.B.~is supported by the DFG Sonderforschungsbereich/Transregio 9.

\begin{appendix}

\section{One-loop corrections to $\phi_\pi(u)$}   

\label{app:phipi}

We briefly summarize our results for the individual diagrams in
Figure~\ref{fig:relcorr} in the light meson case (in Feynman gauge). For
simplicity we present the results for the leading-twist LCDA
$\phi_\pi(u)$ and stick to the case $m_1=m_2=m$.   

\subsection{Vertex diagram}

Starting from the NR limit $\phi_\pi(u)=\delta(u-1/2)$ and performing
the loop-integral in $D=4-2\epsilon$ dimensions, one obtains for the
first diagram in Figure~\ref{fig:relcorr} the distribution  
\begin{eqnarray}
  I_a(u) &\propto&
4 \, \Gamma(\epsilon)  
\left( 
  \frac{\mu^2 e^{\gamma_E}}{m^2 \, (1-2u)^2} \right)^\epsilon   
\left(
  1- \epsilon \, \frac{4 u^2 - 4u -1}{(1-2u)^2} \right)
\left[
  u \, \theta(1-2u)+ \bar u  \,\theta(2u-1) \right] \,.
\end{eqnarray} 
The integral contains an UV-divergence reflected by
$\Gamma(\epsilon)$. The IR-divergence at $u=1/2$ can be isolated with
the help of a plus-distribution which we introduce via
(\ref{eq:plusplusdef}). With this we obtain     
\begin{eqnarray}
I_a(u) &\propto&
4 \left[
\left( \frac{1}{\epsilon} + \ln \frac{\mu^2}{m^2 (1-2u)^2} -
  \frac{4u^2-4u-1}{(1-2u)^2} \right) 
\left[ u \, \theta(1-2u) + \bar u \, \theta(2u-1) \right] 
\right]_{++}
\cr
&&
+ \left( \frac{3}{\epsilon} + 3 \, \ln \frac{\mu^2}{m^2} - 2 \right)
\delta(u-1/2).
\end{eqnarray}
Notice that the term with $\delta'(u-1/2)$ vanishes due to the symmetry
$u \leftrightarrow \bar u$ in the equal mass case (the
``++''-distribution actually coincides with the usual ``+''-distribution
in this case). The local term determines a correction to the decay
constant and does not contribute to $\phi_\pi(u)$.

\subsection{Wilson-line diagrams}

For the second diagram in Figure~\ref{fig:relcorr} we obtain
\begin{eqnarray}
I_b(u) &\propto&
8 \, \Gamma(\epsilon)
\, \int_0^{1/2} dv \,  
   \left( \frac{\mu^2 e^{\gamma_E}}{m^2 (1-2v)^2} \right)^\epsilon
   \frac{v}{2v-1} 
\left[ \delta(u-1/2) - \delta(u-v) \right].
\end{eqnarray}
Convoluting with a regular test function, we get
\begin{eqnarray}
\int_0^1 du \,  f(u) \; I_b(u) &\propto&
\int_0^1 du \,  f(u) \; 
\left[ \frac{8 u \theta(1-2u)}{1-2u} \left(   \frac{1}{\epsilon} + \ln
    \frac{\mu^2}{m^2 (1-2u)^2} \right) \right]_{+}.
\end{eqnarray}
The other Wilson-line diagram in Figure~\ref{fig:relcorr} is obtained from
$I_b$ by symmetrization $u \to \bar u$. 

\section{One-loop corrections to $\phi^{\pm}_B(\om)$}

\label{app:phim}

In the following we present our results for the diagrams in
Figure~\ref{fig:relcorr} in the heavy meson case (in Feynman gauge). We
compute the first order corrections to the NR limit at the matching
scale $\mu\sim m$ starting from $\phi_B^{\pm}(\om_{\rm
  in})=\delta(\om_{\rm in}-m)$, and the general anomalous dimension
kernels related to the renormalization of $\phi_B^+(\om;\mu)$ and
$\phi_B^-(\om;\mu)$. The latter are extracted from the UV-divergent
parts of the diagrams, where we consider arbitrary input functions
$\phi_B^\pm(\om_{\rm in})$ and also keep track of the light quark mass
$m\neq 0$. Notice, that a possible mixing of the 3-particle LCDAs into
$\phi_B^-(\omega;\mu)$ is not considered. Some care has to be taken when
performing the collinear limit (which amounts to setting the transverse
momentum of the incoming light antiquark to zero).

\subsection{Vertex diagram}

The loop-integral in the first diagram of Figure~\ref{fig:relcorr} reads 
\begin{eqnarray}
\left( \begin{array}{c} 
I_a^{+}(\omega)
  \\
I_a^{-}(\omega)
 \end{array} \right)
  &\propto& -\int d\om_{\rm in} \,
    \int  [dl] \,
    \frac{\delta(\om-\om_{\rm in} + n_-l) }
     {[v\cdot l + i0]\,[l^2 + i0]\,
      [(l-k)^2 + i0]}
\nonumber \\[0.3em]
&& \ \times
\left(
  \begin{array}{cc}
    n_-l - \om_{\rm in} + 
 \frac{m^2}{\om_{\rm in}} \left( \frac{k_\perp \cdot l_\perp}{k_\perp^2}
   - 1 \right) 
&
 - m \, \frac{k_\perp \cdot l_\perp}{k_\perp^2}
\\
 - m \, \frac{k_\perp \cdot l_\perp}{k_\perp^2}
&
   n_+l - \frac{m^2}{\om_{\rm in}} +
   \omega_{\rm in} \left( \frac{k_\perp \cdot l_\perp}{k_\perp^2} - 1
   \right) 
  \end{array}
\right)
\left( \begin{array}{c}
\phi_B^+(\om_{\rm in})
\\
\phi_B^-(\om_{\rm in})       
       \end{array}
\right) \nonumber\\[-0.5em]
\label{Ia1}
\end{eqnarray}
where $\om_{\rm in} = n_- k$ with $k^\mu$ being the momentum of the
incoming spectator quark and $k^\mu - l^\mu$ is the spectator-quark
momentum after the interaction with the gluon. We also performed the
collinear limit $k_\perp \to 0$, which requires to keep terms of order
$k_\perp \cdot l_\perp/k_\perp^2$.

Let us first consider the fixed-order corrections to the NR limit, where
$\phi_B^\pm(\om_{\rm in})=\delta(\om_{\rm in}-m)$. Then the loop
integrals simplify according to  
\begin{eqnarray}
  I_a^{\pm}(\omega) &\propto& \int [dl] \, 
    \frac{2m - n_\mp l}
     {[v\cdot l + i0]\,[l^2 + i0]\,
      [l^2 - 2m \, v \cdot l  + i0]}
\ \delta(\om-m+n_-l) \,.
\label{ia1}
\end{eqnarray}
Performing the loop-integrals in $D=4-2\epsilon$ dimensions, one is left
with the distributions  
\begin{eqnarray}
  I_a^+(\omega) &\propto&
 2 \om \, \Gamma(1+\epsilon)
 \left( \frac{\mu^2 e^{\gamma_E}}{(m-\omega)^2} \right)^\epsilon   
\left\{ \frac{2}{(m-\om)^2} 
 - \frac{\theta(m-\omega)}{m (m-\om)} - \frac{\theta(\omega-m)}{\om
   (\om-m)} \right\}  
\nonumber\\[0.3em]
&={}&
 4 \om \left[ \frac{\theta(2m-\om)}{(m-\om)^2} \right]_{++}  
+ 4 \om \, \frac{\theta(\om-2m)}{(m-\om)^2}
- 2 \om \left[ \frac{\theta(m-\om)}{m(m-\om)} +
  \frac{\theta(\om-m)}{\om(\om-m)} \right]_+
 \nonumber\\[0.2em]
&& {}  
  + 2 \left( \frac{1}{\epsilon}+\ln \frac{\mu^2}{m^2} -4 \right)
  \delta(\om-m), \\[0.5em]
  I_a^-(\omega) &\propto&
 \frac{2 (1-\epsilon) \, \Gamma(\epsilon)}{m} 
 \left( \frac{\mu^2 e^{\gamma_E}}{(m-\omega)^2} \right)^\epsilon 
 \, \theta(m-\omega) 
\no \cr
&& {} + 2 m \, \Gamma(1+\epsilon)
 \left( \frac{\mu^2 e^{\gamma_E}}{(m-\omega)^2} \right)^\epsilon 
\Big\{ \frac{2}{(m-\om)^2}  -\frac{\theta(m-\omega)}{m(m-\om)}-
\frac{\theta(\omega-m)}{m(\om-m)} \Big\}  
\nonumber\\[0.3em]
&={}& 2 \left( \frac{1}{\epsilon} + \ln \left[\frac{\mu^2}{(m-\om)^2} \right] - 1 \right) \frac{\theta(m-\om)}{m}
 + 4m \left[ \frac{\theta(2m-\om)}{(m-\om)^2}\right]_{++}
 + 4m \,\frac{\theta(\om-2m)}{(m-\om)^2} 
\nonumber\\[0.2em]
&& {} - 2 \left[ \frac{\theta(m-\om)}{m-\om} \right]_+ 
      - 2 \om \left[ \frac{\theta(\om-m)}{\om(\om-m)} \right]_+
+ 2 \left( \frac{1}{\epsilon} + \ln \frac{\mu^2}{m^2} -4 \right)
\delta(\om-m).
\label{Iaminus}
\end{eqnarray} 
The integration over $\om$ determines the local vertex correction to be
absorbed into the decay constant
\beq
\int_0^\infty d\omega \, I_a^\pm(\omega)
&\propto&
\left( \frac{3}{\epsilon} + 3\, \ln \frac{\mu^2}{m^2} - 2 \right).
\eeq

Focusing now on the UV-divergent contributions to the integration
kernels in the general case, we find
\begin{eqnarray}
  I_a^{+}(\omega) \big|_{\rm div.} &\propto& {\cal O}(\epsilon^0) \,,
\\[0.3em]
 I_a^{-}(\omega) \big|_{\rm div.} &\propto& 
 \frac{2}{\epsilon} \, \int d\om_{\rm in} \left(
   \frac{\theta(\om_{\rm in}-\om)}{\om_{\rm in}}  \right)
 \phi_B^-(\omega_{\rm in})   
+ {\cal O}(\epsilon^0)\,.
\end{eqnarray}
Notice that only the kernel in $I_a^-$ receives
an UV-divergent piece, which can be traced back to the appearance
of a factor $(n_+l)$ in the numerator of (\ref{Ia1}). 
The fact that the kernel
of $I_a^+$ is UV-finite is in line 
with the findings of \cite{Lange:2003ff}.

\subsection{Wilson-line coupling to heavy quark}

In this case there is no mixing between $\phi_B^+$ and $\phi_B^-$ since
the light-quark propagator is not involved  
\begin{eqnarray}
 I_b^\pm(\om) &\propto& 
-  \int d\om_{\rm in} \, \int [dl] \, 
    \frac{\delta(\omega-\omega_{\rm in}+n_-l) - \delta(\omega-\omega_{\rm in}) 
          }{(n_-l) \, [v\cdot l + i0]\,[l^2 + i0]} \, \phi_B^{\pm}(\om_{\rm in}) \,. 
\end{eqnarray}
Inserting the non-relativistic LCDAs and performing the $(n_+l)$ and
$l_\perp$ integrations, one is left with the parameter integral
($k=-n_-l$) 
\begin{eqnarray}
  I_b^\pm(\omega) &\propto&
 2 \, \Gamma(\epsilon)
\, \int_0^\infty dk \, 
   \left( \frac{\mu^2 e^{\gamma_E}}{k^2} \right)^\epsilon
 \frac{\delta(\omega-m-k)-\delta(\omega-m)}{k} .
\end{eqnarray}
Notice that the remaining integral induces an additional UV-divergence,
which has to be isolated by introducing appropriate
plus-distributions. We find
\begin{eqnarray}
I_b^\pm(\om)&\propto&
2 \, \omega \left[\left(\frac{1}{\epsilon} + \ln
    \left[\frac{\mu^2}{(\om-m)^2}\right]  \right) 
 \frac{\theta(\om-m)}{\om (\om-m)}\right]_+
-  \left( \frac{1}{\epsilon^2}+ 
\frac{1}{\epsilon} \, \ln \frac{\mu^2}{m^2}
 + \frac12 \ln^2 \frac{\mu^2}{m^2}
 + \frac{3 \pi^2}{4} \right) \delta(\om-m).\nonumber\\[-0.5em]
\end{eqnarray}
The UV-divergent contribution for $I_b^+$ corresponds to the result for
the diagram (D1) in \cite{Lange:2003ff} (with $\omega'\equiv m$). The
UV-divergence from $k\to\infty$ is a peculiarity of the heavy meson wave
function. It is related to the cusp-anomalous dimension involving the
heavy quark (characterized by a time-like vector $v^\mu$) and the soft
Wilson line (characterized by a light-like vector $n_-^\mu$). The
resulting $1/\epsilon^2$ terms are universal for $\phi_B^+$ and
$\phi_B^-$, 
\begin{eqnarray}
  I_b^\pm(\om) \big|_{\rm div.}&\propto& 
- 
\left( \frac{1}{\epsilon^2} + \frac{1}{\epsilon} \, \ln \frac{\mu^2}{\om^2} \right)\, \phi_B^\pm(\om)
+ \frac{2}{\epsilon} \, \int d\om_{\rm in} \,
\left[\frac{\theta(\om-\om_{\rm in})}{\om-\om_{\rm in}} \right]
\left[\phi_B^\pm(\omega_{\rm in}) - \phi_B^\pm(\om) \right]
+
{\cal O}(\epsilon^0) \,.\nonumber
\\[-0.5em]
\end{eqnarray}

\subsection{Wilson-line coupling to light quark}

In this case the loop integrals mix $\phi_B^+$ into $\phi_B^-$ (but not
vice versa), 
\begin{eqnarray}
\left( \begin{array}{c} 
I_c^{+}(\omega)
  \\
I_c^{-}(\omega)
 \end{array} \right)
  &\propto& 2\int d\om_{\rm in} \,
    \int [dl]\,
    \frac{\delta(\omega-\om_{\rm in}+n_-l) - \delta(\omega-\om_{\rm
        in})}{(n_-l) \, [(l-k)^2 + i0] \, [l^2+i0]} 
\nonumber \\[0.3em]
&& \ \times
\left(
  \begin{array}{cc}
    \om_{\rm in} -  n_-l & 0
\\
  m \, \frac{k_\perp \cdot l_\perp}{k_\perp^2}
&
   \omega_{\rm in} \left( 1-\frac{k_\perp \cdot l_\perp}{k_\perp^2} \right)
  \end{array}
\right)
\left( \begin{array}{c}
\phi_B^+(\om_{\rm in})
\\
\phi_B^-(\om_{\rm in})       
       \end{array}
\right) \,.
\label{Ic1}
\end{eqnarray}
Inserting the non-relativistic LCDAs and performing the $(n_+l)$ and
$l_\perp$ integrations, we find ($k=n_-l$)
\begin{eqnarray}
  I_c^+(\omega) &\propto&
 2 \, \Gamma(\epsilon)
\, \int_0^m dk \, \frac{m-k}{m} 
   \left( \frac{\mu^2 e^{\gamma_E}}{k^2} \right)^\epsilon
 \frac{\delta(k-m+\omega)-\delta(\omega-m)}{k}
\nonumber \\[0.2em]
&{}=&
2 \om \left[ \left(\frac{1}{\epsilon} + \ln \left[\frac{\mu^2}{(\om-m)^2}\right]\right)
 \frac{\theta(m-\omega)}{m \, (m-\omega)} \right]_+
+  \left(\frac{2}{\epsilon}+ 2 \ln\frac{\mu^2}{m^2}+4 \right)
\delta(\om-m),
\\[0.5em] 
  I_c^-(\omega) &\propto&
 2 \, \Gamma(\epsilon)
\, \int_0^m dk 
   \left( \frac{\mu^2 e^{\gamma_E}}{k^2} \right)^\epsilon
 \frac{\delta(k-m+\omega)-\delta(\omega-m)}{k}
\nonumber \\[0.2em]
&{}=&
2 \left[ \left(\frac{1}{\epsilon} + \ln \left[\frac{\mu^2}{(\om-m)^2} \right]\right)
\,
 \frac{\theta(m-\omega)}{m-\omega} \right]_+ \,.
\end{eqnarray} 

The UV-divergent contributions to the integration kernels are identified as
\begin{eqnarray}
  I_c^+(\om) \big|_{\rm div.}&\propto& 
\frac{2}{\epsilon} \, \phi_B^+(\om)
+ \frac{2}{\epsilon} \, \int d\om_{\rm in}
\left[\frac{\om\, \theta(\om_{\rm in}-\om)}{\om_{\rm in} \, (\om_{\rm in}-\om)} \right]
 \left[\phi_B^+(\omega_{\rm in}) - \phi_B^+(\omega) \right]
+ {\cal O}(\epsilon^0) \,,
\\[0.5em]
 I_c^-(\om)\big|_{\rm div.} &\propto& \frac{2}{\epsilon}
\, \phi_B^-(\om) +
\frac{2}{\epsilon} \, \int d\om_{\rm in} 
\left[\frac{\om\, \theta(\om_{\rm in}-\om)}{\om_{\rm in} \, (\om_{\rm in}-\om)} \right] \left[  \phi_B^-(\om_{\rm in}) - \phi_B^-(\om)\right]
\nonumber\\[0.2em]
&&
{}\quad +\frac{2}{\epsilon} \, \int d\om_{\rm in}
\left[\frac{m \, \theta (\om_{\rm in}-\om)}{\om_{\rm in}^2}\right] \left[
  \phi_B^+(\om_{\rm in}) - \phi_B^+(\om) \right]
+ {\cal O}(\epsilon^0) \,.
\end{eqnarray}
Our result for $I_c^+$ is in line with \cite{Lange:2003ff}. In
particular, there are no additional UV-divergences related to cusp
anomalous dimensions since the light-quark and the soft Wilson line are
characterized by the same light-cone vector $n_-^\mu$. For a
non-vanishing light-quark mass, the result for $I_c^-$ implies that the
LCDA $\phi_B^+$ mixes into $\phi_B^-$ under evolution. In the massless
case, however, $\phi_B^+$ and $\phi_B^-$ evolve independently (at least
to leading logarithmic approximation).

\section{Equations of motion for heavy meson LCDAs} 

\label{app:eom}

In this appendix we show that the eom-constraint (\ref{eom1}) holds
after including first order relativistic corrections to the NR limit. We
first evaluate the right-hand side of (\ref{eom1}) using our explicit
results for the 3-particle LCDAs from (\ref{3particle}) 
\begin{align}
&  (D-2) \, \int_0^\omega d\eta \int_{\omega-\eta}^\infty
\frac{d\xi}{\xi} \, \frac{\partial}{\partial \xi}
\left[\Psi_A(\eta,\xi)-\Psi_V(\eta,\xi)\right] 
\cr
&\quad\!= {}
\frac{\alpha_s C_F}{4\pi} 
\left\{ \left( \frac{1}{\varepsilon} + \ln \frac{\mu^2}{m^2} \right)  
m \, \delta(\omega-m)
- \left[ \frac{2\om}{m} \left( \frac{1}{\varepsilon} + \ln
     \frac{\mu^2}{(m-\om)^2} +1 \right) + 2  \ln \frac{(m-\om)^2}{m^2}
 \right] \theta(m-\om)\right\}.\nonumber\\[-1em]
\end{align}
For the expressions on the left-hand side of (\ref{eom1}), we obtain
\begin{align}
&\omega \, \phi_B^-(\omega) - m^{\rm bare} \, \phi_B^+(\omega)
\cr
&\qquad= (m^{\rm OS} - m^{\rm bare}) \; \delta(\omega-m)
 + \frac{\alpha_s C_F}{4\pi} \left\{ 
\frac{2\om}{m} \left(\frac{1}{\varepsilon} + \ln \frac{\mu^2}{(m-\om)^2}
  -1 \right) \theta(m-\om) \right.\cr
&\qquad \qquad \left. +
\left( \frac{2}{\varepsilon} +2 \ln\frac{\mu^2}{(\om-m)^2} -2
\right)\theta(\om-m) - \left( \frac{2}{\varepsilon} +2
  \ln\frac{\mu^2}{m^2} +4 \right)  m \, \delta(\omega-m) \right\}.
\end{align}
and
\begin{align}
& \frac{D-2}{2} \, \int_0^\omega d\eta
\left[\phi_B^+(\eta)-\phi_B^-(\eta) \right]  
\cr
&\qquad= {} -\frac{\alpha_s C_F}{4\pi} 
\left\{ \left[ \frac{4\om}{m} \left( \frac{1}{\varepsilon} + \ln
      \frac{\mu^2}{(m-\om)^2} \right) +2  \ln \frac{(m-\om)^2}{m^2}
  \right]  \theta(m-\om) \right.\cr
&\qquad \qquad \left. +
\left(  \frac{2}{\varepsilon} +2 \ln \frac{\mu^2}{(\om-m)^2} -2
\right) \theta(\om-m) \right\}.
\end{align}
Noticing that 
$$
(m^{\rm OS} - m^{\rm bare}) = \frac{\alpha_s C_F}{4\pi} \left(
  \frac{3}{\epsilon} +   3 \, \ln \frac{\mu^2}{m^2} + 4 \right) m +
{\cal O}(\alpha_s^2),
$$
we see that the equation of motion (\ref{eom1}) is indeed fulfilled
after including the $\alpha_s$ corrections.

\section{Solution of RGE for $\phi_B^-(\omega)$ in WW approximation}

\label{app:evol}

We derive the solution of the evolution equation (\ref{evolapprox}) for
the $B$\/-meson LCDA $\phi_B^-(\om;\mu)$, ignoring the possible mixing
with 3-particle LCDAs and neglecting the light quark mass $m$. We follow
the analysis in~\cite{Lee:2005gz}, where a closed form for the LCDA
$\phi_B^+(\om;\mu)$ to LL approximation has been given. When the
3-particle LCDAs are neglected, $\phi_B^-(\om;\mu)$ can be related to
$\phi_B^+(\om;\mu)$ by the Wandzura-Wilczek relation (\ref{ww0}). This
is also reflected in the leading-order result for the anomalous
dimension kernels $\gamma_-^{(1)}(\om,\om';\mu)$ from
(\ref{gammaminuskernel}) and $\gamma_+^{(1)}(\om,\om';\mu)$ from
(\ref{LN-kernel}). Noticing that    
\begin{align}
  - \om \, \frac{d}{d\omega} 
 \, \int_0^\eta \frac{d\om'}{\eta}  
 \left\{
 \left(
   \Gamma_{\rm cusp}^{(1)} \, \ln \frac{\mu}{\om} -2\right) 
      \delta(\om-\om') 
  - \Gamma_{\rm cusp}^{(1)} \, \frac{\theta(\om'-\om)}{\om'}
 \right\} & {} = \left(
   \Gamma_{\rm cusp}^{(1)} \, \ln \frac{\mu}{w} -2\right) \, \delta(\om-\eta)
\,,
\end{align}
and
\begin{align}
  \int_0^\infty d\eta \; f(\eta) \;
  \left\{
     - \om \, \frac{d}{d\omega} 
     \, \int_0^\eta \frac{d\om'}{\eta}
     \left[\frac{\theta(\om-\om')}{\om-\om'} \right]_+ 
   \right\}
  {} = {}  & - \om \, \frac{d}{d\omega} \,
  \int_0^1 \frac{dx}{x} \int_0^x \frac{dy}{1-y} \, f(\om (1-y))
\nonumber \\[0.2em]
  {} = {} \int_0^1 \frac{dx}{x} \left[ f(\om(1-x))- f(\om) \right] 
  {} = {} &\int_0^\infty d\eta \; f(\eta) \;
  \left[\frac{\theta(\om-\eta)}{\om-\eta} \right]_+ \,,
\end{align}
and
\begin{align}
  \int_0^\infty d\eta \; f(\eta) \;
  \left\{
     - \om \, \frac{d}{d\omega} 
     \, \int_0^\eta \frac{d\om'}{\eta}
     \left[\frac{\om \, \theta(\om'-\om)}{\om'\, (\om'-\om)} \right]_+ 
   \right\}
  {} = {}  &  \om \, \frac{d}{d\omega} \,
\int_0^\infty \frac{dx}{x (1+x)} \int_0^x \frac{dy}{1+y} \, f(\om (1+y))
\nonumber \\[0.2em]
  {} = {} \int_0^\infty \frac{dx}{x (1+x)} \left[ f(\om(1+x))- f(\om) \right]
  {} = {} &\int_0^\infty d\eta \; f(\eta) \;
  \left[\frac{\om \, \theta(\eta-\om)}{\eta \,(\eta-\om)} \right]_+ \,,
\end{align}
we find that the anomalous dimensions fulfill the relation
\begin{equation}
  - \om \, \frac{d}{d\om} \, \int_0^\eta \frac{d\om'}{\eta}
  \, \gamma_-^{(1)}(\om,\om';\mu) = \gamma_+^{(1)}(\om,\eta;\mu) \,.
\end{equation}
Therefore, the functions $ \Phi_B^-(\om;\mu) \equiv \om \,
d\phi_B^-(\om;\mu)/d\om$ and $\phi_B^+(\om;\mu)$ obey the same evolution
equation to LL approximation. Using an intermediate result
from~\cite{Lee:2005gz}, we write the solution for $\Phi_B^-(\om;\mu)$ as
\begin{align}
\Phi_B^-(\om;\mu) = {} & 
 e^{V-2\gamma_E \,g}
 \left(\frac{\om}{\mu_0} \right)^g \,
 \int_0^\infty \frac{d\om'}{\om'} \, \Phi_B^-(\om';\mu_0) \;
 \sum_{m=1}^\infty \,
  \frac{(-1)^{m+1} \, \Gamma(1+m-g)}{\Gamma(1-m+g)\Gamma(1+m)\Gamma(m)}
\nonumber \\[0.2em]
& {} \times \left\{ 
  \theta(\om-\om') \, \left(\frac{\om}{\om'}\right)^{-m}
 +\theta(\om'-\om) \, \left(\frac{\om}{\om'}\right)^{m-g} \right\}
\nonumber \\[0.3em]
= {} & 
 e^{V-2\gamma_E \,g}
 \left(\frac{\om}{\mu_0} \right)^g \,
 \int_0^\infty \frac{d\om'}{\om'} \, \phi_B^-(\om';\mu_0) \;
 \sum_{m=1}^\infty \,
  \frac{(-1)^{m} \, \Gamma(1+m-g)}{\Gamma(1-m+g)\Gamma(1+m)\Gamma(m)}
\nonumber \\[0.2em]
& {} \times \left\{ 
 m  \, \theta(\om-\om') \, \left(\frac{\om}{\om'}\right)^{-m}
 + (g-m) \, \theta(\om'-\om) \, \left(\frac{\om}{\om'}\right)^{m-g}
\right\}  \,,
\end{align}
with $V$ and $g$ from (\ref{Vdef},\ref{gdef}) and we assumed $0<g<1$. We
finally perform the $\omega$--integral to obtain $\phi_B^-(\om;\mu)$
from $\Phi_B^-(\om;\mu)$, and the summation over $m$ which leads to
hypergeometric functions with the result,
\begin{align}
 \phi_B^-(\om;\mu)  {} = {} &
 e^{V-2 \, \gamma_E \, g } \,
 \frac{\Gamma(1-g)}{\Gamma(g)} \,
\int_0^\infty
\frac{d\om'}{\om'} \, \phi_B^-(\om';\mu_0) \, \left(\frac{\om}{\mu_0}\right)^g
\cr 
 & \times \left\{
  {}  \theta(\om-\om') \ \frac{\om'}{\om} \
  {}_2F_1(1-g,1-g,1,\om'/\om)  
\right.
\cr 
& \quad \left. 
  {} + \theta(\om'-\om) 
  \left(\frac{\om}{\om'}\right)^{-g}
  \, {}_2F_1(1-g,1-g,1,\om/\om') \right\} \,.
\end{align}

\end{appendix}

%\clearpage

\end{document}